\documentclass[sigconf]{acmart}
\usepackage{booktabs} 
\newcommand{\eat}[1]{}
\usepackage{multirow}
\usepackage{graphicx}
\usepackage{subfigure}
\usepackage{caption}
\usepackage{pifont}
\usepackage{float}
\usepackage{enumitem}

\usepackage{fancyhdr}
\usepackage{xcolor}

\AtBeginDocument{%
  \providecommand\BibTeX{{%
    \normalfont B\kern-0.5em{\scshape i\kern-0.25em b}\kern-0.8em\TeX}}}

\copyrightyear{2020}

\begin{document}

\title{\includegraphics[width=\columnwidth]{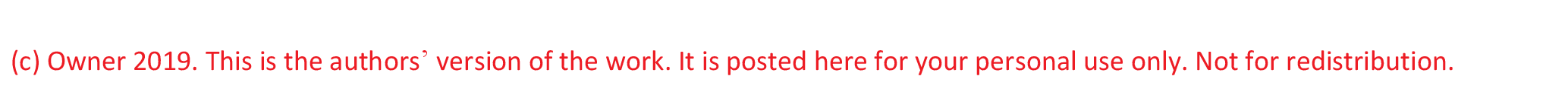} A Synopses Data Engine for Interactive Extreme-Scale Analytics}
\titlenote{This work has received funding from the EU Horizon 2020 research
and innovation program INFORE under grant agreement No 825070.}
 
\author{Antonis Kontaxakis}
\affiliation{%
  \institution{Athena Research Center \\ Technical University of Crete}
}
\email{akontaxakis@athenarc.gr}
\email{akondaxakis@softnet.tuc.gr}

\author{Nikos Giatrakos}
\affiliation{%
  \institution{Athena Research Center \\ Technical University of Crete}
}
\email{ngiatrakos@athenarc.gr}
\email{ngiatrakos@softnet.tuc.gr}

\author{Antonios Deligiannakis}
\affiliation{%
  \institution{Athena Research Center  \\ Technical University of Crete}
}
\email{adeli@athenarc.gr}
\email{adeli@softnet.tuc.gr}

\renewcommand{\topfraction}{0.9}
\renewcommand{\dbltopfraction}{0.9}
\renewcommand{\floatpagefraction}{.8}
\renewcommand{\floatsep}{6pt}
\renewcommand{\dblfloatsep}{6pt}
\renewcommand{\textfloatsep}{6pt}
\renewcommand{\dbltextfloatsep}{6pt}

\begin{abstract}
In this work, we detail the design and structure of a Synopses Data Engine (SDE)
which combines the virtues of parallel processing and stream summarization
towards delivering interactive analytics at extreme scale. Our SDE is built on top of
Apache Flink and implements a synopsis-as-a-service paradigm. In that it achieves
(a) concurrently maintaining thousands of synopses of various types for thousands of streams on demand,
(b) reusing maintained synopses among various concurrent workflows,
(c) providing data summarization facilities even for cross-(Big Data) platform workflows,
(d) pluggability of new synopses on-the-fly,
(e) increased potential for workflow execution optimization. The proposed SDE is useful for interactive analytics at extreme scales
because it enables (i) enhanced horizontal scalability, i.e., not only scaling out the computation to a number of processing units
available in a computer cluster,
but also harnessing the processing load assigned to each by operating on carefully-crafted data summaries, (ii) vertical
scalability, i.e., scaling the computation to very high numbers of processed streams and (iii) federated scalability i.e.,
scaling the computation beyond single clusters and clouds by
controlling the communication required to answer global queries posed over a number of potentially geo-dispersed clusters.
\end{abstract}

%
%


\maketitle

\section{Introduction}\label{sec:intro}

Interactive extreme-scale analytics over massive, high speed data streams become of the essence in a wide variety of modern
application scenarios. In the financial domain, NYSE alone generates several terrabytes of data a day, including trades of thousands of stocks~\cite{forbes}. Stakeholders such as authorities and investors need to analyze these data in an interactive, online fashion for timely market surveillance or investment risk/opportunity identification purposes. In the life sciences domain, studying the effect of applying combinations of drugs on simulated tumors of realistic sizes can generate cell state data of 100 GB/min~\cite{DBLP:journals/tasm/GiatrakosKDAGPA19}, which need to be analyzed online to interactively determine successive drug combinations. In maritime surveillance applications, one needs to fuse high-velocity position data streams of hundreds of thousands of vessels across the globe and satellite, aerial images~\cite{Bereta2019} of various resolutions. In all these scenarios, data volumes and rates are only expected to rise in the near future. In the financial domain, data from emerging markets, such as crypto-currencies, are increasingly added to existing data sources. In life sciences, simulations are becoming progressively more complex, involving billions of interacting cells, while in the maritime domain
autonomous vehicles are added as on-site sensing information sources.

To enable interactive analytics at extreme-scale, stream processing platforms and systems need to provide
three types of scalability:
\begin{itemize}[leftmargin=*]

\item Horizontal scalability, i.e., the ability to scale the computation with extreme data volumes and
data arrival rates as analyzed in the aforementioned scenarios. This requires scaling out the computation to a number of machines and respective processing units available at a corporate data center (cluster) or cloud. Horizontal scalability is achieved by parallelizing the processing and adaptively assigning computing resources to running analytics queries. 

\item Vertical scalability, i.e., the ability to scale the computation with the number of processed streams. For instance, to detect systemic risks in the financial scenario, i.e., stock level events that could trigger instability or collapse of an entire industry or economy, requires discovering and interactively digging into correlations among tens of thousands of stock streams. The problem involves identifying the highly correlated pairs of stock data streams under various statistical measures, such as Pearson's correlation over $N$ distinct, high speed data streams, where $N$ is a very large number. To track the full $\Theta(N^2)$ correlation matrix results in a quadratic explosion in space and computational complexity which is simply infeasible for very large $N$. The problem is further exacerbated when considering higher-order statistics (e.g., conditional dependencies/correlations). The same issue arises in the maritime surveillance scenario for trajectory similarity scores over hundreds of thousands of vessels. Clearly, techniques that can provide vertical scaling are sorely needed for such scenarios. 

\item Federated scalability, i.e., the ability to scale the computation in settings where data arrive at multiple, potentially
geographically dispersed sites. On the one hand, a number of benchmarks~\cite{DBLP:journals/pvldb/ZeitlerR11, DBLP:conf/icde/KarimovRKSHM18} conclude that, in such settings, even if horizontal scalability is ensured within each cluster, the maximum achieved throughput (number of streaming tuples that are processed per time unit) is network bound. On the other hand, consider again the systemic risk detection scenario from the financial domain where stock trade data arrive at geo-dispersed data centers around the globe. Moving entire data streams around the sites in order to extract pairwise correlation scores depletes the available bandwidth, introducing network latencies that prevent the interactivity of the desired analytics.  
\end{itemize}

Big Data platforms, including Apache Flink~\cite{flink}, Spark~\cite{spark}, Storm~\cite{storm} among others,
have been developed that support or are especially dedicated to stream processing. Such platforms
focus on horizontal scalability, but they are not sufficient by themselves to allow for the required vertical and federated scalability.
On the other hand, there is a wide consensus in stream processing~\cite{DBLP:conf/vldb/ZhuS02, DBLP:conf/vldb/MankuM02, DBLP:journals/jal/CormodeM05, Flajolet07hyperloglog:the, DBLP:journals/ftdb/CormodeGHJ12,
DBLP:books/sp/16/GarofalakisGR16, CormodeYi} that approximate but rapid answers to analytics tasks, more often than not, suffice.
For instance, knowing in real-time that a group of approximately 50 stocks, extracted out of thousands or millions of stock combinations, is highly (e.g., $>0.9$ score) correlated is more than sufficient to detect systemic risks. Therefore, such an approximate result is preferable
compared to an exact but late answer which says that the actual group is composed of 55 stocks with correlation scores accurate to the last decimal. Data summarization techniques such as samples, sketches or histograms~\cite{DBLP:journals/ftdb/CormodeGHJ12} build carefully-crafted synopses of Big streaming Data which preserve data properties important for providing approximate answers, with tunable accuracy guarantees, to a wide range of analytic queries. Such queries include, but are not limited to, cardinality, frequency moment, correlation, set membership or quantile estimation~\cite{DBLP:journals/ftdb/CormodeGHJ12}. 

Data synopses enhance the horizontal scalability provided by Big Data platforms. This is because parallel versions of
data summarization techniques, besides scaling out the computation to a number of processing units, reduce the volume
of processed high speed data streams. Hence, the complexity of the problem at hand is harnessed
and execution demanding tasks are severely sped up. For instance, sketch summaries~\cite{DBLP:journals/tods/CormodeG08} can aid in tracking
the pairwise correlation of streams in space/time that is sublinear in the size of the original streams. Additionally, data synopses
enable vertical scalability in ways that are not possible otherwise. Indicatively, the coefficients of Discrete Fourier
Transform(DFT)-based synopses~\cite{DBLP:conf/vldb/ZhuS02} or the number of set bits (a.k.a. Hamming Weight) in Locality Sensitive Hashing(LSH)-based bitmaps~\cite{DBLP:journals/is/GiatrakosKDVT13} have been used for correlation-aware hashing of streams to respective processing units. Based on the synopses, using DFT coefficients or Hamming Weights as the hash key respectively,
highly uncorrelated streams are assigned to be processed for pairwise comparisons at different processing units. Thus, such
comparisons are pruned for streams that do not end up together. Finally, federated scalability is ensured both by the fact
that communication is reduced since compact data stream summaries are exchanged among the available sites and by exploiting the mergeability property~\cite{Agarwal:2012:MS:2213556.2213562} of many synopses techniques. As an example, answering cardinality estimation queries over a number of
sites, each maintaining its own FM sketch~\cite{DBLP:journals/ftdb/CormodeGHJ12} is as simple as communicating only small bitmaps (typically 64-128 bits) to the
query source and performing a bitwise \texttt{OR} operation.

In this work, we detail the design and structure of a Synopses Data Engine (SDE) built on top of
Apache Flink ingesting streams via Apache Kafka~\cite{kafka}. Our SDE combines the virtues of parallel processing and stream summarization
towards delivering interactive analytics at extreme scale by enabling enhanced horizontal, vertical and federated
scalability as described above. However, the proposed SDE goes beyond that. Our design implements a Synopsis-as-a-Service (termed SDEaaS) paradigm where the SDE can serve multiple, concurrent application workflows in which each maintained synopsis can be used as an operator. That is, our SDE operates as a single, constantly running Flink job which achieves:
\begin{enumerate}[label=\Alph*.,leftmargin=*]
\item concurrently maintaining thousands of synopses for thousands of streams on demand,
\item reusing maintained synopses among multiple application workflows (submitted jobs) instead of redefining and duplicating streams
for each distinct workflow separately,
\item pluggability of new synopses' definitions on-the-fly,
\item providing data summarization facilities even for cross-(Big Data) platform workflows~\cite{DBLP:conf/icde/KaoudiQ18} outside of Flink,
\item optimization of workflows execution by enabling clever data partitioning,
\item advanced optimization capabilities to minimize workflow execution times by replacing exact operators (aggregations, joins etc) with approximate ones, given a query accuracy budget to be spent.
\end{enumerate}

Few prior efforts provide libraries for online synopses maintenance, but neglect parallelization aspects~\cite{stream-lib,DataSketch}, or lack a SDEaaS design~\cite{proteus} needing to run a separate job for each maintained synopsis. The latter compromises aspects in points A-F above and increases cluster scheduling complexity. Others~\cite{DBLP:reference/bdt/Mozafari19} lack architectural provisions for federated scalability and are limited to serving simple aggregation operators being deprived of vertical scalability features as well.
On the contrary, our proposed SDE not only includes provisions for federated scalability and provides a rich library of synopses to be loaded and maintained on the fly, but also allows
to plug-in external, new synopsis definitions customizing the SDE to application field needs. More precisely, our contributions are:
\begin{enumerate}[leftmargin=*]
\item We present the novel architecture of a Synopses Data Engine (SDE) capable of providing interactivity in extreme-scale analytics
by enabling various types of scalability.  
\item Our SDE is built using a SDE-as-a-Service (SDEaaS) paradigm, it can efficiently maintain thousands of synopses for thousands of streams to serve multiple, concurrent, even cross-(Big Data) platform, workflows.
\item We describe the structure and contents of our SDE Library, the implemented arsenal including data summarization techniques for the proposed SDE, which is easily extensible by exploiting inheritance and polymorphism.
\item We discuss insights we gained while materializing a SDEaaS paradigm and outline lessons learned useful for future, similar endeavors.
\item We showcase how the proposed SDE can be used in workflows to serve a variety of purposes towards achieving interactive data analytics.
\item We present a detailed experimental analysis using real data from the financial domain to prove the ability of our approach to scale
at extreme volumes, high number of streams and degrees of geo-distribution, compared to other candidate approaches. 
\end{enumerate}

\section{Related Work}\label{sec:related}
From a research viewpoint, there is a large number of related works on data synopsis techniques.
Such prominent techniques, cited in Table~\ref{tab:library}, are already incorporated in our SDE
and, some of them, are further discussed in practical examples in Section~\ref{sec:workflows}.
Please refer to~\cite{DBLP:journals/ftdb/CormodeGHJ12,
DBLP:books/sp/16/GarofalakisGR16, CormodeYi} for comprehensive views on relevant issues.  

Yahoo!DataSketch~\cite{DataSketch}
and Stream-lib~\cite{stream-lib} are software libraries of stochastic streaming algorithms and
summarization techniques, correspondingly. These libraries are detached from parallelization
and distributed execution aspects, contrary to the SDE we propose in this work. Apache Spark~\cite{spark}, provides
utilities for data synopsis via sampling operators, CountMin sketches and Bloom Filters. Similarly, Proteus~\cite{proteus}
extends Flink with data summarization utilities.
Spark and Proteus combine the potential of data summarization with parallel processing over Big Data platforms
by providing libraries of data synopsis techniques. Compared to these, first, we provide a richer library of data summarization techniques (Table~\ref{tab:library}) which covers all types of scalability mentioned in
Section~\ref{sec:intro}. Second, we propose a novel architecture for implementing a SDEaaS paradigm which allows synopses to be loaded from internal libraries or get plugged from external libraries on-the-fly, as the service is up and running. Third,
as we detail in Section~\ref{sec:lessons}, our SDEaaS paradigm and architecture enable the simultaneous maintenance of thousands of synopses of different types for thousands of streams which (i) might not even be possible without our SDEaaS paradigm, (ii) allows various running workflows to share/reuse currently maintained synopses and thus prevents duplicating the same data and synopses for each workflow, (iii) reduces
the load of cluster managers compared to accomplishing the same task, but lacking our SDEaaS design. Finally, SnappyData's~\cite{DBLP:reference/bdt/Mozafari19} stream processing is based on Spark. SnappyData's SDE is limited to serving simple \texttt{SUM}, \texttt{COUNT} and \texttt{AVG} queries\footnote{
https://snappydatainc.github.io/snappydata/aqp/\#overview-of-synopsis-data-engine-sde} being deprived of vertical scalability features
and federated scalability architectural provisions.

\section{SDE API -- Supported Operations}\label{sec:operations}

\begin{figure*}[t]
\centering
\begin{minipage}[b]{.25\textwidth}
     \centering\includegraphics[width=0.65\columnwidth]{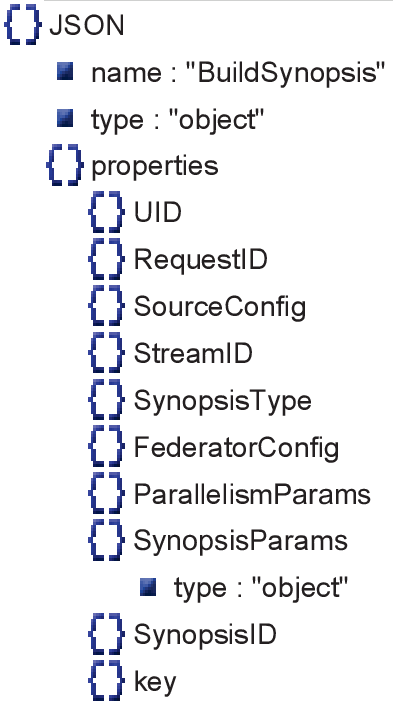}
    \caption{JSON snippet for \texttt{BuildSynopsis} request.}
    \label{fig:json}
\end{minipage} \qquad
\begin{minipage}[b]{0.7\textwidth}
  \centering\includegraphics[width=\textwidth]{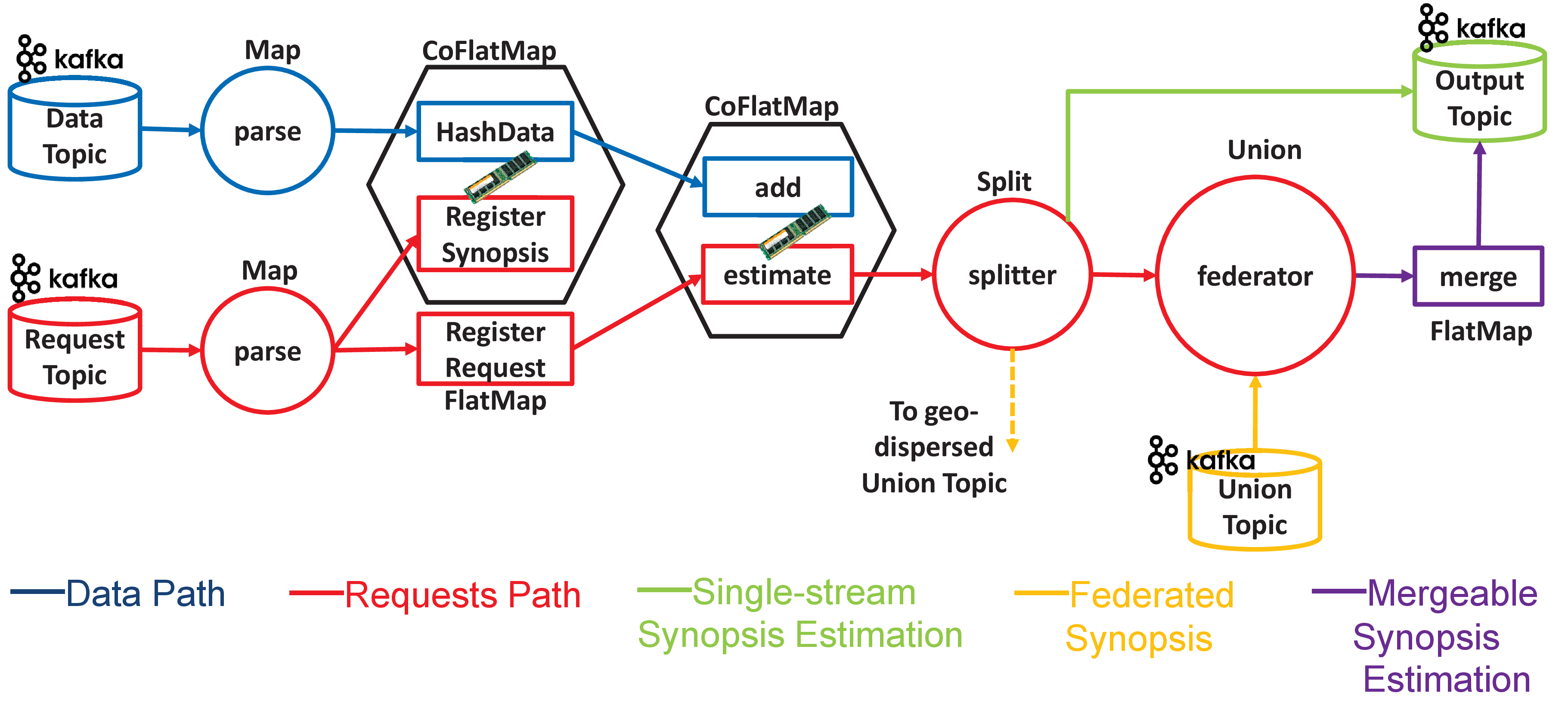}
    \caption{SDE Architecture -- Condensed View.}
    \label{fig:architecture}
\end{minipage}
\end{figure*}

In this section, we outline the functionality that our SDE API provides to upstream
(i.e., contributing input to) and downstream (receiving input from)
operators and application interfaces of a given Big Data processing
pipeline engaging synopses. All requests are submitted to the SDE at runtime, given the SDEaaS nature of our design, via lightweight, properly formatted JSON snippets~\cite{YahooBenchMark} to ensure cross-(Big Data) platform compatibility. The JSON snippet of each request listed below includes a unique identifier for the queried stream or source incorporating multiple streams (see Section~\ref{sec:architecture}) and a unique id for the synopsis to be
loaded/created/queried. In case of a create or load synopsis request (see below and Table~\ref{tab:library}), the parameters of the synopsis as well as a pair of parameters involving the employed parallelization degree and scheme (see Section~\ref{sec:architecture})
are also included in the JSON snippet. In federated architectures where multiple, geo-dispersed clusters run local SDEaaS instances and estimations provided by synopses need to be collected at a cluster afterwards, the JSON snippet also includes the address of that cluster. The SDE API provides the following facilities:\\
\noindent{\bf \texttt{Build/Stop Synopsis} (Request)\/}.
A synopsis can be created or ceased on-the-fly, as the SDE is up and running. In that, the execution of other running workflows that
utilize synopsis operators, is not hindered. A synopsis may be (a) a single-stream synopsis, i.e., a synopsis (e.g. sample) maintained on the trades of a single stock, or (b) a data source synopsis, i.e., a synopsis maintained on all trades irrespectively of the stock. Moreover, \texttt{Build/Stop Synopsis} allows submitting a single request for maintaining a synopsis of the same kind, for each
out of multiple streams coming from a certain source. For instance, maintaining a sample per stock for thousands of stocks coming from the same source requires the submission of a single request. A condensed view of a JSON snippet for building a new synopsis is illustrated in
Figure~\ref{fig:json}.\\ 
\noindent{\bf \texttt{Load Synopsis} (Request)\/}. 
The SDE Library (Section~\ref{sec:library}) incorporates a number of synopsis operators, commonly used in practical scenarios.
\texttt{Load Synopsis} supports pluggability of the code of additional (not included in the SDE Library) synopses, their dynamic loading and maintenance at runtime. The structure of the SDE Library, utilizing inheritance and polymorphism, is key for this task. This is an important feature
since it enables customizing the SDE to application specific synopses without stopping the service.\\
\noindent{\bf \texttt{Ad-hoc Query} (Request)\/}.
The SDE accepts one-shot, ad-hoc queries on a certain synopsis and provides respective estimations (approximate answers) to downstream operators or application interfaces, based on its current status. \\
\noindent{\bf \texttt{Continuous Querying}\/}.
Continuous queries can be defined together with a \texttt{Build/Stop Synopsis} request. In this case, an estimation of the approximated
quantities, such as counts, frequency moments or correlations are provided every time the estimation of the synopsis is updated, for instance, due to
reception of a new tuple. 

The response to ad-hoc or continuous queries is also provided in lightweight JSON snippets including: (i) a key, value pair for uniquely identifying the provided response (e.g. from past and future ones) and for the value of the estimated quantity, respectively, (ii) the id of the request that generated the response, (iii) the identifier of the utilized synopses along with its
parameters (Table~\ref{tab:library}).\\
\noindent{\bf \texttt{SDE Status Report}\/}.
The API allows querying the SDE about its status, returning information about the currently maintained synopses and their parameters.
This facility is useful during the definition of new workflows, since it allows each application to discover whether it can utilize already maintained data synopses and reuse synopses serving multiple workflows.

\section{SDE Architecture}\label{sec:architecture}

In this section we detail the SDE architectural components and present their utility in serving the operations specified in Section~\ref{sec:operations}. 

\subsection{SDE Fundamentals}\label{sec:fundamentals}
Our architecture is built on top of Apache Flink~\cite{flink} and Kafka~\cite{kafka}. Kafka is used as a fast, scalable, durable, and fault-tolerant publish-subscribe messaging system enabling connectivity between the SDE and upstream, downstream operators in the workflows served by the SDE. Kafka together with the JSON format of accepted request snippets allows us to materialize the SDEaaS paradigm even when upstream or downstream operators run on different Big Data platforms. Furthermore, it is used as a messaging service in case of querying synopses maintained at a number of geo-dispersed clusters. A Kafka cluster is composed of a number of brokers, run in parallel, that handle separate partitions of topics. Topics constitute categories of data where producers and consumers can write and read, respectively. In the case of the SDE,
producers constitute upstream operators, while downstream operators act as consumers. Furthermore, in geo-dispersed, multi-cluster
settings, the SDE instances run at each cluster may be the producers or consumers of a particular Kafka topic as will be explained later on in this section.

A Flink cluster is composed of (at least one) Master and a number of Worker nodes. The
Master node runs a JobManager for distributed execution and coordination purposes, while each Worker node incorporates
a TaskManager which undertakes the physical execution of tasks. Each Worker (JVM process) has a number of task slots
(at least one).  Each Flink operator may run in a number of instances, executing the same code, but on different data partitions.
Each such instance of a Flink operator is assigned to a slot
and tasks of the same slot have access to isolated memory shared only among tasks of that slot.
Figure~\ref{fig:architecture} provides a condensed view of the SDE architecture, which engages \texttt{Map}, \texttt{FlatMap}, \texttt{CoFlatMap}, \texttt{Union} and \texttt{Split} Flink operators. In a nutshell, a \texttt{Map} operator takes one tuple and produces another tuple in the output, a \texttt{FlatMap} operator takes one tuple and produces zero, one, or more tuples, while a \texttt{CoFlatMap} operator hosts two \texttt{FlatMap} that share access to common variables (therefore the linking icon in the figure) among streams that have previously been connected (using a \texttt{Connect} operator in Flink). Finally, a \texttt{Union} operator
receives two or more streams and creates a new one containing all their elements, while a
\texttt{Split} operator splits the stream into two or more streams according to some criterion.

Section~\ref{sec:components} explains the reason for the above design and explains the flow of information in different uses of the SDE. 
  
\subsection{SDE Architectural Components}\label{sec:components}

\noindent{\bf Employed Parallelization Scheme(s)\/}. The parallelization scheme that is employed in the design of the SDE is partition-based parallelization~\cite{DBLP:journals/vldb/GiatrakosAADG20}. That is, every data tuple that streams in the SDE architecture and is destined to be included in a maintained synopsis, does so based on the partition key it is assigned to it. When a synopsis is maintained for a particular stream (i.e., per stock - see Section~\ref{sec:operations}) the key that is assigned to the respective update (newly arrived data tuple) is the identifier of that particular stream for which the synopsis is maintained. In this case, within the distributed computation framework of Flink, that stream is processed by a task of the same worker and parallelization is achieved by distributing the number of streams for which a synopsis is built, to the available workers in the cluster hosting the SDE. On the other hand, when a synopsis involves a data source (i.e., financial data source for all monitored stock streams  - see Section~\ref{sec:operations}), the desired degree of parallelism is included as a parameter in the respective request to build/start maintaining the synopsis. In the latter case, one dataset is partitioned to the available workers in a round-robin fashion and the respective keys are created by the SDE (details on that follow shortly) each of which points (is hashed) to a particular worker. Finally, in case of processing streaming windows (either tuple or
count-based)~\cite{DBLP:journals/vldb/GiatrakosAADG20} an incoming tuple may (i) initiate a new window, (ii) be assigned to one or more existing windows or (iii) terminate a window. Here, the partition is the window itself and the tuple is given the key(s) of the window(s) it affects.  

\noindent{\bf Data and Query Ingestion\/}. 
Data and request (JSON snippet) streams arrive at a particular Kafka topic each. In the case of the \texttt{DataTopic} of Figure~\ref{fig:architecture}, a parser component is used in order to extract the key and value field(s) on which a currently running synopsis is maintained. The respective parser of the \texttt{RequestTopic} topic of Figure~\ref{fig:architecture} reads the JSON snippet of the request and processes it. When an incoming request involves the maintenance of a new synopsis, the parser component extracts information about the parameters of the synopsis
(see Table~\ref{tab:library}) and its nature, i.e. whether it is on a single stream, on a data source, if it involves a multi-stream synopsis maintenance request or a synopsis that is also maintained in SDE instances in other geo-dispersed clusters. In case the request is an ad-hoc query the parser component extracts the corresponding synopsis identifier(s).

\noindent{\bf Requesting New Synopsis Maintenance\/}.
When a request is issued for maintaining a new synopsis, it initially follows the red-colored paths of the SDE architecture in Figure~\ref{fig:architecture}. That is, the corresponding parser sends the request to a \texttt{FlatMap} operator(termed \texttt{RegisterRequest} at the bottom of Figure~\ref{fig:architecture}) and to another \texttt{FlatMap} operator (\texttt{RegisterSynopsis}) which is part of a \texttt{CoFlatMap} one.  \texttt{RegisterRequest} and \texttt{RegisterSynopsis} produce the keys (as analyzed in the description of the supported parallelization schemes) for the maintained synopsis, but provide different functionality. The \texttt{RegisterRequest} operator uses these keys in order to later decide which worker(s) an ah-hoc query, which also follows the red-colored path, as explained shortly, should reach. On the other hand, the \texttt{RegisterSynopsis} operator uses the same keys to decide to which worker(s) a data tuple destined to update one or more synopses, which follows the blue-colored path in Figure~\ref{fig:architecture}, should be directed. The possible parallelization degree of the \texttt{RegisterSynopsis} and \texttt{RegisterRequest} operators up to this point of the architecture depends on the number of running synopses.

\noindent{\bf Updating the Synopsis\/}.
When a data tuple destined to update one or more synopses is ingested via the \texttt{DataTopic} of Kafka it follows the blue-colored path of the SDE architecture in Figure~\ref{fig:architecture}. The tuple is directed to the \texttt{HashData} \texttt{FlatMap} of the corresponding \texttt{CoFlatMap} where the keys (stream identifier for single stream synopsis and/or worker identifier for data source synopsis and windowing operations) are looked up based on what \texttt{RegisterSynopsis} has created. Following the blue-colored path, the tuple is directed to a \texttt{add} \texttt{FlatMap} operator which is part of another \texttt{CoFlatMap}. The \texttt{add} operator updates the maintained synopsis as prescribed by the algorithm of the corresponding technique. For instance, in case a FM sketch~\cite{DBLP:journals/jcss/FlajoletM85} is maintained, the \texttt{add} operation hashes the incoming tuple to a position of the maintained bitmap and turns the corresponding bit to 1 if it is not already set. 

\noindent{\bf Ad-hoc Query Answering\/}.
 An ad-hoc query arrives via the \texttt{Reque}-\texttt{stTopic} of Kafka and is directed to the \texttt{RegisterRequest} operator. The operator which has produced the keys using the same code as \texttt{RegisterSynopsis} does, looks up the key(s) of the queried synopsis and directs the corresponding request to the \texttt{estimate} \texttt{FlatMap} operator of the corresponding \texttt{CoFlatMap}. The  \texttt{estimate} operator reads via the shared state the current status of the maintained synopsis and extracts the estimation of the corresponding quantity the synopsis is destined to provide. For instance, upon performing an ad-hoc query on a FM sketch~\cite{DBLP:journals/jcss/FlajoletM85}, the  \texttt{estimate}  operator reads the maintained bitmap, finds the lowest position of the unset bit and provides a distinct count estimation by using the index of that position and a $\phi=0.77$ coefficient. Table~\ref{tab:library} summarizes the estimated quantities each of the currently supported synopses can provide.
 
\noindent{\bf Continuous Query Answering\/}.
In case continuous queries are to be executed on the maintained synopses, a new estimation needs to be provided every time the estimation of the synopsis is updated, either via an \texttt{add} operation or because a window on the data expires. In this particular occasion \texttt{estimate} needs to be invoked by \texttt{add}. 

Both in ad-hoc and continuous querying, the result of \texttt{estimate}, following the red path in Figure~\ref{fig:architecture}, is directed to a \texttt{Split} operator, termed \texttt{splitter}. If necessary, the \texttt{splitter} forwards estimations to a \texttt{Union} operator, termed \texttt{federator} which reads from a \texttt{Union} Kafka topic (yellow path in Figure~\ref{fig:architecture}). The \texttt{Union} Kafka topic and the \texttt{federator} involve our provisions for maintaining federated synopses, i.e., synopses that are kept at a number of potentially geo-dispersed clusters. The \texttt{splitter} distinguishes between
three cases. {\bf Case 1:} Case 1 happens when \texttt{estimate} involves a single-stream synopsis maintained only locally at a cluster. Then, \texttt{Split} directs the output to downstream operators of the executed workflow via Kafka, by following the green-colored path in Figure~\ref{fig:architecture}. {\bf Case 2:} Case 2 arises when a federated synopsis is queried but the request has identified
another cluster responsible for extracting the overall estimation. Then, \texttt{Split} acts as the producer (writes) to the 
geo-dispersed \texttt{Union} Kafka topic of another cluster (declared by the dotted, yellow arrow coming out of \texttt{splitter} in 
Figure~\ref{fig:architecture}). Let us now see {\bf Case 3:} For non-federated synopses defined on entire data sources (e.g., a sample over all stock data), a number of workers of the current cluster participate in the employed parallelization scheme as discussed at the beginning of Section~\ref{sec:components}. Thus, each such worker provides its local synopsis/estimation. Because something similar holds when individual clusters maintain federated synopses and the current cluster is set as responsible for synthesizing the overall estimation, in both cases the output of the \texttt{Split} operator is directed via \texttt{Union} to a \texttt{merge} \texttt{FlatMap} following the purple-colored path. The \texttt{merge} operator merges the partial results of the various workers and/or clusters and produces the final estimation which is streamed to downstream operators, again via an \texttt{Output} Kafka topic. For instance, FM sketches~\cite{DBLP:journals/jcss/FlajoletM85} or Bloom Filters~\cite{DBLP:journals/cacm/Bloom70} (bitmaps) can be merged via simple logical disjunctions or conjunctions. At this point, in order to direct all partial estimates to the same worker of a cluster to perform the \texttt{merge} operation, a corresponding identifier for the issued request (for ad-hoc queries) or an identifier for the maintained synopsis (for continuous queries) is used as the key.

\begin{table}
\centering
\resizebox{\columnwidth}{!}{%
\begin{footnotesize}
\begin{tabular}{|c|c|c|}
\hline 
Synopsis & Estimation & Parameters\tabularnewline
\hline 
\hline 
CountMin~\cite{DBLP:journals/jal/CormodeM05} & Count/Frequency Estimation & $\epsilon$,$\delta$\tabularnewline
\hline 
BloomFliter~\cite{DBLP:journals/cacm/Bloom70} & Set Membership & \#elements, False Positive Rate\tabularnewline
\hline 
FM Sketch~\cite{DBLP:journals/jcss/FlajoletM85} & Distinct Count & Bitmap size, $\epsilon$, $\delta$\tabularnewline
\hline 
HyperLogLog~\cite{Flajolet07hyperloglog:the} & Distinct Count & Relative Standard Error\tabularnewline
\hline 
\multirow{2}{*}{AMS Sketch~\cite{DBLP:conf/stoc/AlonMS96}} & $L_{2}$-Norm,  & \multirow{2}{*}{$\epsilon$, $\delta$}\tabularnewline
 & Inner Product &\tabularnewline
\hline 
Discrete Fourier  & \multirow{2}{*}{Correlation, BucketID} & Similarity Threshold,\tabularnewline
Transform (DFT)~\cite{DBLP:conf/vldb/ZhuS02} &  & Number of Coefficients\tabularnewline
\hline
Random Hyperplane &  \multirow{2}{*}{Correlation, BucketID} & Bitmap Size, Similarity Threshold \tabularnewline
Projection (RHP)~\cite{DBLP:conf/stoc/Charikar02,DBLP:journals/is/GiatrakosKDVT13} &  & Number of Buckets \tabularnewline
\hline 
Lossy Counting~\cite{DBLP:conf/vldb/MankuM02} & Count, Frequent Items & $\epsilon$\tabularnewline
\hline 
Sticky Sampling~\cite{DBLP:conf/vldb/MankuM02} & Count, Frequent Items & support, $\epsilon$, $\delta$\tabularnewline
\hline 
Chain Sampler~\cite{DBLP:conf/soda/BabcockDM02} & Sample & Sample Size\tabularnewline
\hline 
GKQuantiles~\cite{DBLP:conf/sigmod/GreenwaldK01} & Quantiles & $\epsilon$\tabularnewline
\hline 
CoreSetTree~\cite{DBLP:journals/jea/AckermannMRSLS12} & CoreSets & Bucket size, dimensionality\tabularnewline
\hline
\end{tabular}
\end{footnotesize}
}

\caption{Supported synopses. $\epsilon$ is the approximation error bound.
$\delta$ is the probability of failing
to achieve $\epsilon$ accuracy. For synopses that can be maintained over a window, respective
parameters for window definition are added.}\label{tab:library}
\end{table}

\begin{figure}
    \centering\includegraphics[width=\columnwidth]{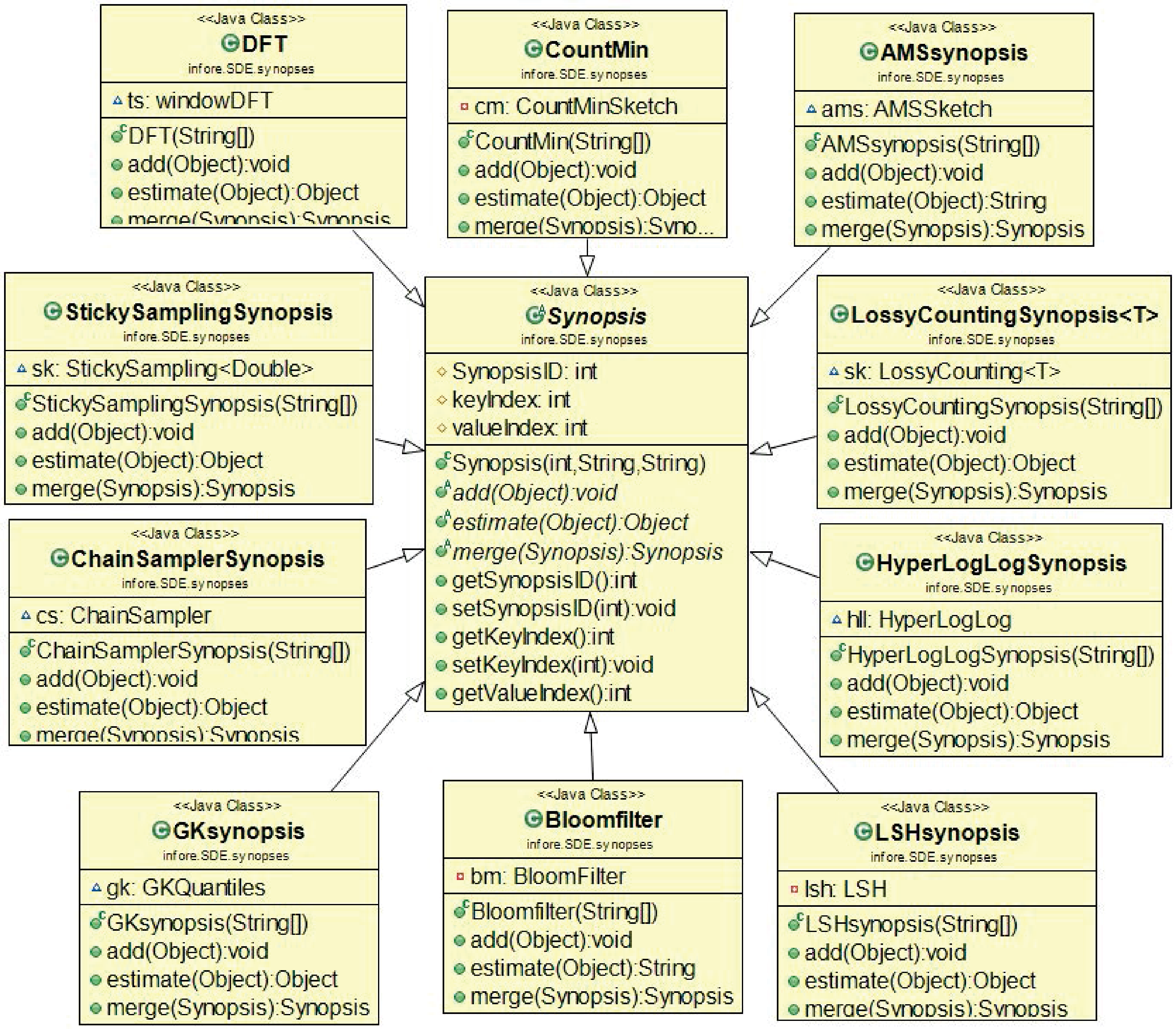}
    \caption{Structure of the Synopses Library (partial view).}
    \label{fig:library}
\end{figure}

\section{SDE Library}\label{sec:library}

The internal structure of the synopses library is illustrated in Figure~\ref{fig:library} which provides only a partial view of the currently supported synopses for readability purposes. Table~\ref{tab:library} provides a full list of currently supported synopses, their utility in terms of estimated quantities and their parameters. The development of the SDE Library exploits subtype polymorphism in Java
in order to ensure the desired level of pluggability for new synopses definitions.

As shown in Figure~\ref{fig:library}, there is a higher level class called \texttt{Synopsis} with attributes related to a unique identifier and a couple of strings. The first string holds the details of the request (JSON snippet) with respect to how the synopsis should be physically implemented, i.e., index of the key field in an incoming data tuple (for single stream synopsis), the respective index of the value field which the summary is built on, whether the synopsis is a federated one and which cluster should synthesize the overall estimation and so on. The second string holds information included in the JSON snippet regarding synopsis parameters as those cited in Table~\ref{tab:library}. Furthermore, the \texttt{Synopsis} class includes methods for \texttt{add}, \texttt{estimate} and \texttt{merge} as those were described in Section~\ref{sec:architecture}. Finally, a set of setters and getters for synopsis, key and value identifiers are provided. 

Every specific synopsis algorithm is implemented in a separate class, as shown in Figure~\ref{fig:library}, that extends \texttt{Synopsis} and overrides the \texttt{add}, \texttt{estimate} and \texttt{merge} methods with the algorithmic details of that particular technique~\cite{DBLP:journals/ftdb/CormodeGHJ12}.  

\section{Insights and Lessons Learned}\label{sec:lessons}
\noindent{\bf Why Flink\/}. In principle, our architectural design can be materialized over other Big Data platforms
such as Storm, Spark or Kafka Streams. The key reason for choosing Flink as the platform for
a proof-of-concept implementation of the proposed architecture is the \texttt{CoFlatMap} operator (transformation).
As shown in the description of our architecture, the fact that \texttt{CoFlatMap} allows two \texttt{FlatMap} operators
gain access to shared variables was used both for generating keys and assign data to partitions processed by certain workers
(leftmost \texttt{CoFlatMap} in Figure~\ref{fig:architecture}) as well as for querying maintained synopses via the \texttt{estimate}
\texttt{FlatMap} in the middle of the figure. Although one can implement the \texttt{CoFlatMap} functionality in
other Big Data platforms, the native support provided by Flink alleviates the development effort with respect to memory configuration, state management and fault tolerance.   

\noindent{\bf The Red Path\/}. Notice, that the blue-colored path in Figure~\ref{fig:architecture} remains totally detached from the red-colored path. This depicts a design choice we follow for facilitating querying capabilities. That is, since the data updates on several maintained synopses may be ingested at an extremely high rate in Kafka at the beginning of the blue path, typically a lot higher than the rate at which requests are issued in the red path, in case the two paths were crossing, back-pressure on the blue-colored path would also affect the timely answers to requests. Having kept the two paths independent, requests can be answered in a timely manner based on the current status of the maintained synopses. This is also true for continuous queries since they
may be interpreted to a number of requests.   

\noindent{\bf ...And One SDEaaS For All\/}. Our SDEaaS approach allows the concurrent maintenance of thousands of synopses for thousands
of streams on demand. It further allows different application workflows to share and reuse existing synopses instead of redefining them.
The alternative is to submit a separate job for each (one or more) desired synopsis in a respective workflow that uses it. The latter simplistic approach
possesses a number of drawbacks. First, the same synopses, even with the exact same parameters, cannot be reused/shared among currently running workflows.  This means that data streams need to be duplicated and redundant data summaries are built as well. Second, one may end up submitting a different job for each new
demand for a maintained synopsis. Apart from increasing the load of a cluster manager, this poses restrictions on the number of synopses that can be simultaneously maintained. Recall from Section~\ref{sec:fundamentals} that each worker in a Flink cluster is assigned a number of task slots and each task slot can host tasks only of the same job. Therefore, lacking our SDEaaS approach means that
the number of concurrently maintained synopses is at most equal to the available task slots. As a rule-of-thumb~\cite{flink}, a default number of task slots would be the number of available CPU cores. Thus, unless thousands of cores are available
one cannot maintain thousands of synopses for thousands of streams. Even when thousands of CPU cores are available, the number of tasks that can run in the same task slot is a multiple of the number of such slots. This observation is utilized in our SDEaaS architecture.
Roughly speaking, in SDEaaS a request for a new synopsis on the fly assigns new tasks for it, while lacking the SDEaaS rationale assigns at least one entire task slot. In SDEaaS, synopses maintenance
involves tasks running instances of the operators in Figure~\ref{fig:architecture}, instead of devoting entire task slots to each. Each synopsis by design consumes limited memory and entails simple update (\texttt{add} in Figure~\ref{fig:architecture}) operations. Thus, in SDEaaS, we have multiple, light tasks virtually competing for task slot resources and better exploit the potential for hyper-threading and pseudo-parallelism for the maintained synopses. For the above reasons, SDEaaS is a much more preferable design choice.   

\noindent{\bf Kafka Topics\/}. In Figure~\ref{fig:architecture} we use four specific Kafka topics which the SDE consumes (\texttt{DataTopic}, \texttt{RequestTopic}, \texttt{UnionTopic}) or produces (\texttt{OutputTopic}, \texttt{UnionTopic}). Our SDE is provided as a service and constantly runs as a single Flink job (per cluster, in federated settings). Synopses are created
and respective sources of data are added on demand, but our experience in developing the proposed SDE says that there is no reliable way of adding/removing new Kafka topics to a Flink job dynamically, at runtime. Therefore all data tuples, requests and outputs need to be written/read
in the respective data topics, each of which may include a number of partitions, i.e. per stream or data source. This by no means introduces redundancy in the data/requests processed by the SDE, because every data tuple that arrives in 
the \texttt{DataTopic} has no reason of existing there unless it updates one or more maintained synopses. Similarly every request that arrives in the \texttt{RequestTopic} creates/queries specific synopses. The same holds for the \texttt{OutputTopic} and \texttt{UnionTopic}. No output is provided unless a continuous query has been defined for a created synopses or an ad-hoc request arrives.
In both cases, the output is meant to be consumed by respective application workflows.
Furthermore, internal to the SDE, nothing is consumed or produced in the \texttt{UnionTopic} unless one or more federated synopses are maintained.    

\noindent{\bf Windows \& Out-of-order Arrival Handling\/}. In Flink, Spark and other Big Data platforms, should a
window operator need to be applied on a stream, one would use a programming syntax similar to (\texttt{\{streamName||operatorName\}.chosenWindowOperator}). If one does that
in a SDEaaS architecture, the window would be applied to the entire operator, i.e., \texttt{CoFlatMap}, \texttt{FlatMap} and so on in
Figure~\ref{fig:architecture}. But, in the general case, each maintained synopsis incorporates the definition of its own window which may differ across different currently maintained synopses, instead of the same window operator applied to all synopses. Therefore, a
SDEaaS design does not allow for using the native windowing support provided by the Big Data platform because the various windows are not known in advance. One should develop custom
code and exploit low-level stream processing concepts provided by the corresponding platform (such as the \texttt{ProcessFunction}
in Flink~\cite{flink}) to implement the desired window functionality. The same holds for handling out-of-order tuple arrivals and
the functionality provided by \texttt{.allowedLateness()} in Flink or similar operators in other platforms.

\noindent{\bf Dynamic Class Loading\/}. YARN-like cluster managers, upon being run as sessions, start the TaskManager and JobManager processes with the Flink framework classes in the Java classpath. Then job classes are loaded dynamically when the jobs are submitted. But what we require in a \texttt{Load Synopsis} request provided by our API is different. Due to the SDEaaS nature of the SDE,
to materialize \texttt{Load Synopsis} we need to achieve loading classes dynamically {\it after} the SDE job has been submitted, as
the service is up and running. A cluster manager will not permit loading classes at runtime due to security issues, i.e. class loaders are to be immutable. In order to bypass such issues for classes involving synopses that are external to our SDE Library, one needs to store the corresponding \texttt{jar} file in HDFS and create an own, child class loader. That is, the child class loader must have a constructor accepting a class loader, which must be set as its parent. The constructor will be called on JVM startup and the real system class loader will be passed. We leave testing \texttt{Load Synopsis} using alternative ways (e.g. via REST API), for future work.  

\section{SDE$\lowercase{aa}$S \& How it can be used}\label{sec:workflows}

\begin{figure}[t]
  \centering\includegraphics[width=.9\columnwidth]{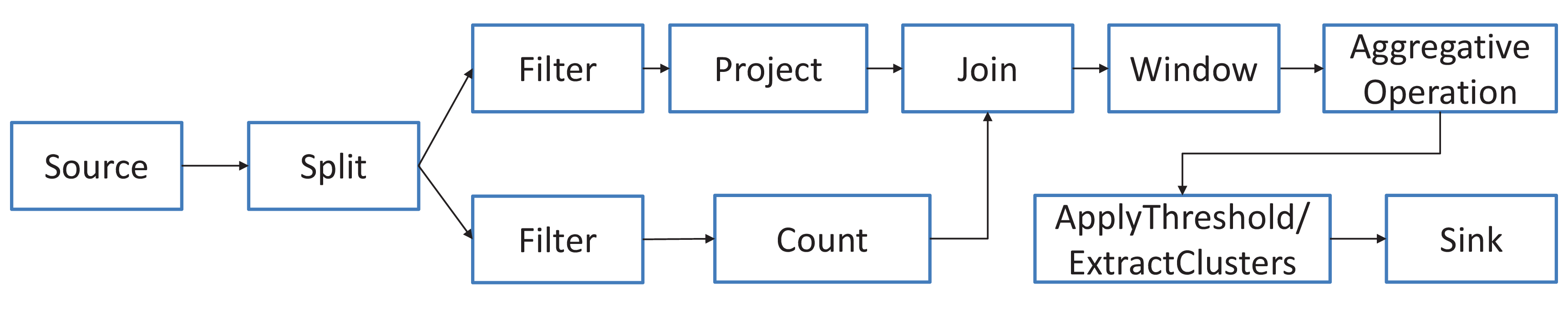}
    \caption{SDEaaS in Practice -- Workflow under Study.}
    \label{fig:workflow}
\end{figure}

In this section we design a specific scenario, we build a workflow that resembles, but extends, the Yahoo! Benchmark~\cite{YahooBenchMark} and then, we discuss how our SDE and its SDEaaS characteristics can be utilized so as to
serve a variety of purposes. Consider our running example from the financial domain. The workflow of Figure~\ref{fig:workflow}
illustrates a scenario that utilizes Level 1 and Level 2 stock data aiming at discovering cross-correlations among and groups of
correlated stocks. More precisely, Level 1 data involve stock trades of the form $<Date, Time, Price, Volume>$ for each data tick of an asset (stock). Level 2 data show the activity that takes place before a trade is made.
Such an activity includes information about offers of shares and corresponding prices as well as respective bids and prices per stock.
Thus, Level 2 data are shaped like series of $<Ask\;price, Ask\;volume, Bid\;price, Bid\;volume>$ until a trade is made. These pairs
are timestamped by the time the stock trade happens. The higher the number of such pairs for a stock, the higher the popularity of the stock. Note that, in Figure~\ref{fig:workflow}, we use generic operator namings. The workflow may be specified in any Big Data platform, other than Flink, and
still use (in ways that we describe here) the benefits of SDEaaS acting as producer (issuing requests) and consumer to the Kafka topics
of Figure~\ref{fig:architecture}, abiding by the respective JSON schemata.  

In Figure~\ref{fig:workflow} both Level 1 and Level 2 data arrive at a \texttt{Source}. The \texttt{Split} operator separates Level 1 from Level 2 data.
It directs Level 2 data to the bottom branch of the workflow. There, the bids are \texttt{Filter}ed (i.e., for monitoring only a subset of stocks or keep only bids above a price/volume threshold). Then, the bids are \texttt{Count}ed and only this counter is kept per stock. When a trade for a stock
is realized, the corresponding Level 1 tuple is directed by \texttt{Split} to the upper part of the workflow. A \texttt{Project} operator keeps only
the timestamp of the trade for each stock. The \texttt{Join} operator afterwards joins the stock trade, Level 1 tuple with the count of bids the stock received until the trade. The
corresponding result is inserted in a time \texttt{Window} of recent such counts, forming a time series. The pairwise similarities of the time series or coresets~\cite{DBLP:journals/jea/AckermannMRSLS12} of stocks are computed via an \texttt{AggregativeOperation}. The results either in the form of pairs of stocks surpassing a similarity threshold (\texttt{ApplyThreshold} operator in Figure~\ref{fig:workflow}) or clusters of stocks (\texttt{ExtractClusters} operator in Figure~\ref{fig:workflow}) are directed to a \texttt{Sink} to support relevant decision making procedures.

\noindent{\bf ...as a Cost Estimator for Enhanced Horizontal Scalability\/}. SDEaaS can act as a cost estimator that constantly collects statistics for streams (in this scenario, stocks) that are of interest and these statistics can be used for optimizing the execution of any currently running or new
workflow. In our examined scenario, having designed the workflow in Figure~\ref{fig:workflow} we wish to determine an appropriate number of workers that will be assigned
for its execution, prescribing the parallelization degree, as well as balance the processing load among the dedicated workers.
For that purpose a HyperLogLog~\cite{Flajolet07hyperloglog:the} and a CountMin~\cite{DBLP:journals/jal/CormodeM05} sketch (see Table~\ref{tab:library}) can be used, i.e., our SDE constantly runs as a service and keeps HLL and CountMin sketches.

HyperLogLog (HLL) sketches~\cite{Flajolet07hyperloglog:the} enable the extraction of approximate distinct counts using limited memory and a simple error approximation formula. Therefore, they are useful for estimating the cardinality of the set of stocks that are being monitored per time unit. In the common implementation of HyperLogLog, each incoming element is hashed to a 64-bit bitmap. The hash function is designed so that the hashed values closely resemble a uniform model of randomness, i.e., bits of hashed values are assumed to be independent and to have an equal probability of occurring each. The first $m$ bits of the bitmap are used for bucketizing an incoming element and we have an array $M$ of $2^m$ buckets (also called registers). The rest $64-m$ bits are used so as to count the number of leading zeros and in each bucket we store the maximum such number of leading zeros to that particular bucket. To extract a distinct count estimation, one needs to compute the harmonic mean of the values of the buckets. The relative error of HLL in the estimation of the distinct count is $1/\sqrt{2^m}$. HLL are trivial to merge based on equivalent number of buckets maintained independently at each site/cluster. One should simply derive the maximum among the corresponding buckets of sites.  

A CountMin Sketch~\cite{DBLP:journals/jal/CormodeM05} is a two dimensional array of $w \times d$ dimensionality used to estimate frequencies of elements of a stream using limited amount of memory. For given accuracy $\epsilon$ and error probability $\delta$,
$w=e/\epsilon$ ($e$ is the Eurler's number)  and $d=log (1/\delta)$. $d$ random, pairwise independent hash functions are chosen for hashing each tuple (concerning a particular stock) to a column in the sketch. When a tuple streams in, it goes through the $d$ hash functions so that one counter in each row is incremented. The estimated frequency for any item is the minimum of the values of its associated counters. This provides an estimation within $\epsilon N$, when $N$ is the sum of all frequencies so far (in the financial dataset), with probability at least $1-\delta$. CountMin sketches are easily mergeable by adding up the corresponding arrays.

An intrinsic optimizer can use SDEaaS as the cost estimator, derive the cardinality of the set of stocks that need to be monitored per time unit by querying the HLL sketch.
Moreover, the CountMin sketch can be queried for estimating the frequency of each stock. Based on the HLL estimation
the optimizer knows how many pieces of work need to be assigned to the workers. And based on the frequency of each stock, the size of each piece of work is also known. Therefore, the optimizer can configure the number of workers and balance the load among them, for instance, by using a Worst Fit Decreasing Bin-packing approach~\cite{DBLP:journals/vldb/GiatrakosAADG20}. Horizontal scalability is
enhanced compared to what is provided by the Big Data platform alone. This is due to having apriori (provided by the SDEaaS nature of the engine) adequate statistics to ensure that no worker is overloaded causing reduction in the overall throughput during the execution of the workflow.

\noindent{\bf ...for Locality-aware Hashing \& Vertical Scalability\/}. Consider that the \texttt{AggregativeOperation} in
Figure~\ref{fig:workflow} involves computing pairwise similarities of stock bid count time series based on Pearson's Correlation
Coefficient. As discussed in Section~\ref{sec:intro}, tracking the full correlation matrix results in a quadratic explosion in space and time which is simply infeasible for very large number of monitored stocks. Let us now see how the DFT synopsis
(Table~\ref{tab:library}) can be used for performing locality-aware hashing of streams to buckets, assign buckets including time series of stocks to workers and prune the number of pairwise comparisons for time series that are not hashed nearby. For that
purpose, the SDE should be queried in-between the \texttt{Window} and \texttt{AggregativeOperation} of Figure~\ref{fig:workflow} so as to get
the bucketID per stock, i.e., the id of the worker where the \texttt{AggregativeOperation} (pairwise similarity estimation) will be performed independently.

Our Discrete Fourier Transform (DFT)-based correlation estimation implementation is based on StatStream~\cite{DBLP:conf/vldb/ZhuS02}. An important  observation
for assigning time series to buckets is that there is a direct relation between Pearson's correlation coefficient (denoted $Corr$ below) among time series $x$, $y$ and the Euclidean distance of their corresponding normalized version (we use primes to distinguish DFT coefficients of normalized time series from the ones of the unnormalized version). In particular, $Corr(x,y)$ $=$
$1-\frac{1}{2} d^2 (X^{\prime},Y^{\prime})$, where $d(.)$ is the Euclidean distance.

The DFT transforms a sequence of $n$ (potentially complex) numbers $x_{0}\ldots,x_{n-1}$ into another sequence of complex numbers $X_{0},\ldots,$ $X_{n-1}$, which is defined by the DFT coefficients, calculated as
$X_F=\frac{1}{n} \sum_{k=1}^{(n-1)}x_k  e^{\frac{i2ðkF}{n}}$, for $F=0,\ldots,n-1$ and $i=\sqrt{-1}$.

Compression is achieved by restricting $F$ in the above formula to few coefficients. There are a couple of additional properties of the DFT which are taken into consideration for parallelizing the processing load of pairwise comparisons among time series:
\begin{enumerate}[leftmargin=*]
\item The Euclidean distance of the original time series and their DFT is preserved. We use this property to estimate the Euclidean distance of the original time series using their DFTs.
\item It holds that $Corr(x,y)$ $\geq$ $1-\epsilon^2$ $\Rightarrow $ $d(X^{\prime},Y^{\prime})\leq \epsilon$. This says that it is meaningful to examine only pairs of time series for which $d(X^{\prime},Y^{\prime})\leq \epsilon$ . We use this property to bucketize (hash) time series based on the values of their first coefficient(s) and then assign the load of pairwise comparisons within each bucket to workers. 
\end{enumerate}
The DFT coefficients can be updated incrementally upon operating over sliding windows~\cite{DBLP:conf/vldb/ZhuS02}. Let us now explain how the time series that are approximated by the DFT coefficients are bucketized so that possibly similar time series are hashed to the same or neighboring buckets, while the rest are hashed to distant buckets and, therefore, they are never compared for similarity. Time series that are hashed to more than one buckets are replicated an equal amount of times. 

Now, assume a user-defined threshold $T$. According to our above discussion, in order for the correlation to be greater than $T$, then $d(X^{\prime},Y^{\prime})$ needs to be lower than $\epsilon$, with $T = 1 - \epsilon^2$. By using the DFT on normalized series, the original series are also mapped into a bounded feature space. The norm (the size of the vector composed of the real and the imaginary part of the complex number) of each such coefficient is bounded by $\sqrt{2} /2$. 

Based on the above observation,~\cite{DBLP:conf/vldb/ZhuS02} notices that the range of each DFT coefficient is between $-\sqrt{2} /2$ and $\sqrt{2} /2$. Therefore, the DFT feature space is a cube of diameter $\sqrt{2}$. Based on this, we use a number of DFT coefficients to define a grid structure, composed of buckets for hashing groups of time series to each of them. Each bucket in the grid is of diameter $\epsilon$ and there are in total $2\lceil \frac{\sqrt{2}}{2\epsilon}\rceil^{(\#used\_coefficients)} $ buckets. For instance,
in~\cite{DBLP:conf/vldb/ZhuS02} 16-40 DFT coefficients are used to approximate stock exchange time series.

Each time series is hashed to a specific bucket inside the grid. Suppose $X^{\prime}$ is hashed to a bucket. To detect the time series whose correlation with $X^{\prime}$ is above $T$, only time series hashed to the same or adjacent buckets are possible candidates. Those time series are a super-set of the true set of highly-correlated ones. Since the bucket diameter is $\epsilon$, time series mapped to non-adjacent buckets possess a Euclidean distance greater than $\epsilon$, hence, their respective correlation is guaranteed to be lower than $T$. Moreover, due to that property, there will be no similarity checks that are pruned while their score would pass the threshold. 

Again, note that here the principal role of the SDEaaS is to produce the corresponding DFT coefficients and hash time series to buckets.
That is why it should be queried between the \texttt{Window} and the \texttt{AggregativeOperation}. Therefore, the output of the corresponding synopsis in Table~\ref{tab:library} includes the resulted coefficients and the bucket identifier. The actual similarity tests (in each bucket) may be performed by the  downstream operator (\texttt{AggregativeOperation}) using the original time series. 

\noindent{\bf ...Synopsis-based Optimization for Enhanced Horizontal Scalability\/}. When an application is willing to bargain accuracy for a considerable processing speed up or reduced memory consumption, the SDEaaS can act as the main tool of an advanced optimizer which would receive the application's accuracy budget and rewrite the workflow to equivalent but approximate forms so as to achieve the aforementioned performance goals. 

Consider the workflow of Figure~\ref{fig:workflow}. Since CountMin sketches are not preferable for correlation
estimation~\cite{DBLP:journals/jal/CormodeM05} in our discussion we are going to engage AMS sketches~\cite{DBLP:conf/stoc/AlonMS96}.
The key idea in AMS sketches is to represent a streaming (frequency) vector $v$ using a much smaller sketch vector $sk(v)$ that is
updated with the streaming tuples and provide probabilistic guarantees for the quality of the data approximation. The AMS sketch defines
the $i$-th sketch entry for the vector $v$, $sk(v)[i]$ as the random variable $\sum_k v[k]\cdot \xi_i [k]$, where $\{\xi_i\}$ is a family
of four-wise independent binary random variables uniformly distributed in $\{-1, +1\}$ (with mutually-independent families across
different entries of the sketch). Using appropriate pseudo-random hash functions, each such family can be efficiently constructed
on-line in logarithmic space. Note that, by construction, each entry of $sk(v)$ is essentially a randomized linear projection (i.e., an
inner product) of the $v$ vector (using the corresponding $\xi$ family), that can be easily maintained (using a simple counter) over the
input update stream. Every time a new stream element arrives, $v[k]\cdot \xi_i [k]$ is added to the aforementioned sum and similarly for
element deletion. Each sketch vector can be viewed as a two-dimensional $w \times d$ array, where $w=O(1/\epsilon^2 )$ and
$d=O(log (1/\delta))$, with $\epsilon$, $1 - \delta$ being the desired bounds on error and probabilistic confidence, correspondingly. The
inner product in the sketch-vector space and the $L_2$ norms (in which case we replace $sk(v_2)$ with
$sk(v_1)$ in the formula below and vice versa) is defined as: $sk(v_{1})\cdot sk(v_{2})=\text{{\scriptsize\ensuremath{\underset{j=1..d}{\underbrace{median}}}}}\left\{ \frac{1}{w}\sum_{i=1}^{w}sk(v_{1})[i,j]\cdot sk(v_{2})[i,j]\right\} $.

Some workflow execution plans that can be produced using our SDEaaS functionality and an accuracy budget are:
\begin{enumerate}[label=Plan\arabic*, leftmargin=*]
\item The \texttt{Count} operator in Figure~\ref{fig:workflow} can be rewritten to a \texttt{SDE.AMS} (sketches) operator
provided by our SDEaaS and then use these sketches to judge pairwise similarities in \texttt{Aggregative} \texttt{Operation}.
\item The \texttt{SDE.DFT} synopsis can replace the \texttt{Window} and \texttt{Aggre} \texttt{gativeOperation} operators to: (i) bucketize time series comparisons, (ii) speed up similarity tests by approximating original time series with few DFT coefficients. 
\item Rewrite the \texttt{Count} operator to \texttt{SDE.AMS} and rewrite the \texttt{Window} and \texttt{Aggre} \texttt{gativeOperation} to \texttt{SDE.DFT} in which case the DFT operates on the sketched instead of the original time series.   
\end{enumerate}
Based on which plans abide by the accuracy budget and on the time and space complexity guarantees of each synopsis, the optimizer can
pick the workflow execution plan that is expected to provide the higher throughput or lower memory usage. Again, horizontal scalability is enhanced compared to what the Big Data platform alone provides, by using the potential of synopses.

\noindent{\bf ...for AQP \& Federated Scalability\/}. In the scope of Approximate Query Processing (AQP), the workflow of Figure~\ref{fig:workflow}
can take advantage of federated synopses that are supported by our SDEaaS architecture (Figure~\ref{fig:architecture}, Section~\ref{sec:components})
in order to reduce the amount of data that are communicated and thus enable federated scalability. For instance,
assume Level 1, Level 2 data of stocks that are being analyzed first arrive at sites (computer clusters each running our SDEaaS) located at the
various countries of the corresponding stock markets. Should one wish to pinpoint correlations of stocks globally, a need
to communicate the windowed time series of Figure~\ref{fig:workflow} occurs. To ensure federated scalability to geo-dispersed
settings composed of many sites, few coefficients of \texttt{SDE.DFT} or \texttt{SDE.AMS} sketches can be used to replace the \texttt{Window} operator in Figure~\ref{fig:workflow} and reduce the dimensionality of the time series. Hence, the communication cost while compressed time series are exchanged among the sites is harnessed and network latencies are prevented.

\noindent{\bf ...for Data Stream Mining\/}. StreamKM++~\cite{DBLP:journals/jea/AckermannMRSLS12} computes a weighted sample of a data stream, called the CoreSet of the data stream. A data structure termed CoreSetTree is used to speed up the time necessary for sampling non-uniformly during CoreSet maintenance. After the CoreSet is extracted from the data stream, a weighted $k$-means algorithm is applied on the CoreSet to get the final clusters for the original stream data. Due to space constraints here, please refer to~\cite{DBLP:journals/jea/AckermannMRSLS12} for further details. In this case, the \texttt{AggregativeOperation} is to be
replaced by \texttt{SDE.CoreSetTree} and the \texttt{ExtractClusters} operator is a weighted $k$-means that uses the CoreSets.

\section{Experimental Evaluation}\label{sec:experiments}

\begin{figure*}[h]
\subfigure[Varying the Parallelism.]{\label{fig:parallelism}\includegraphics[width=.245\textwidth ]{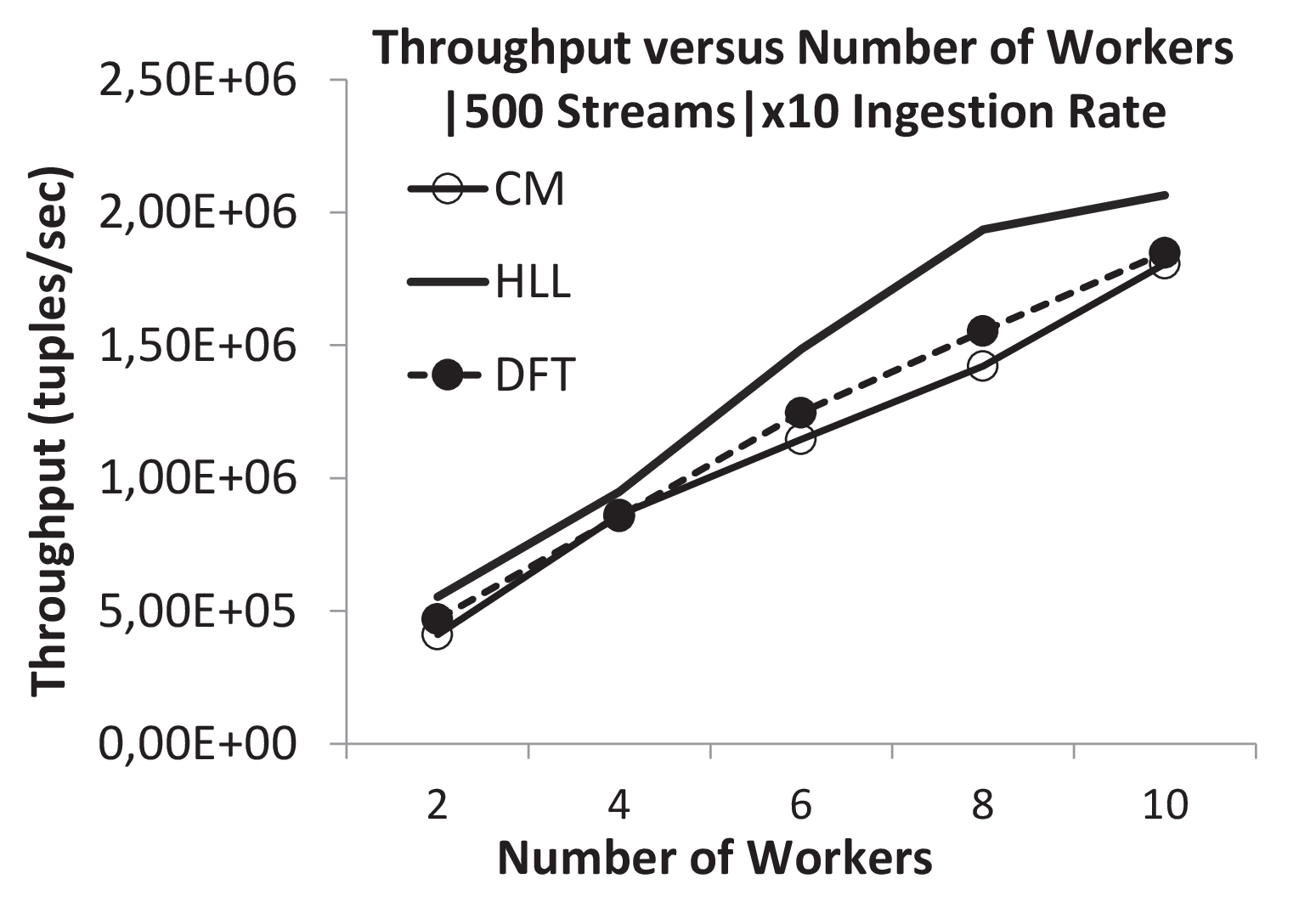}}
\subfigure[Varying the Ingestion Rate.]{\label{fig:injection}\includegraphics[width=.245\textwidth ]{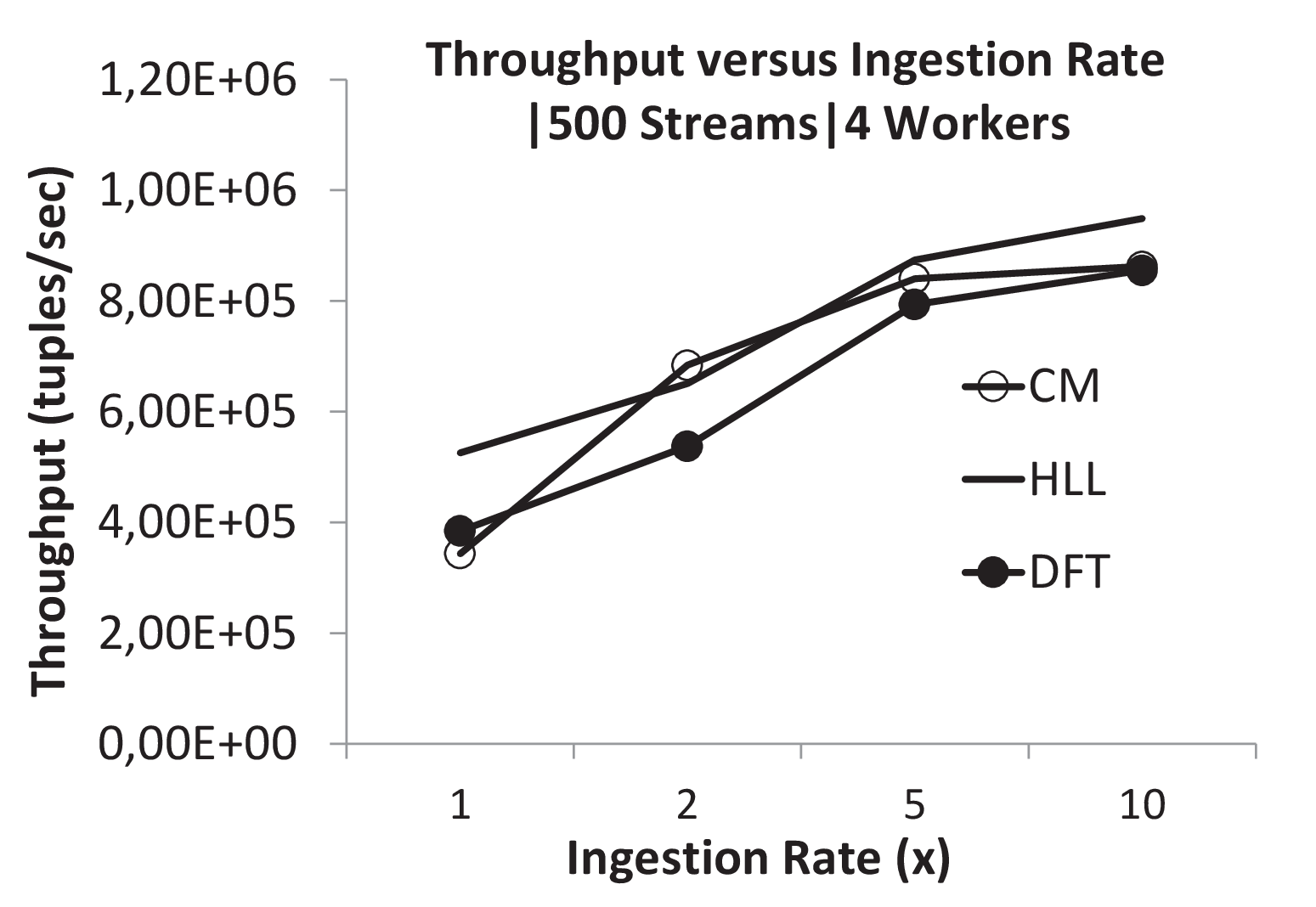}}
\subfigure[Varying the Number of Streams.]{\label{fig:sources}\includegraphics[width=.245\textwidth ]{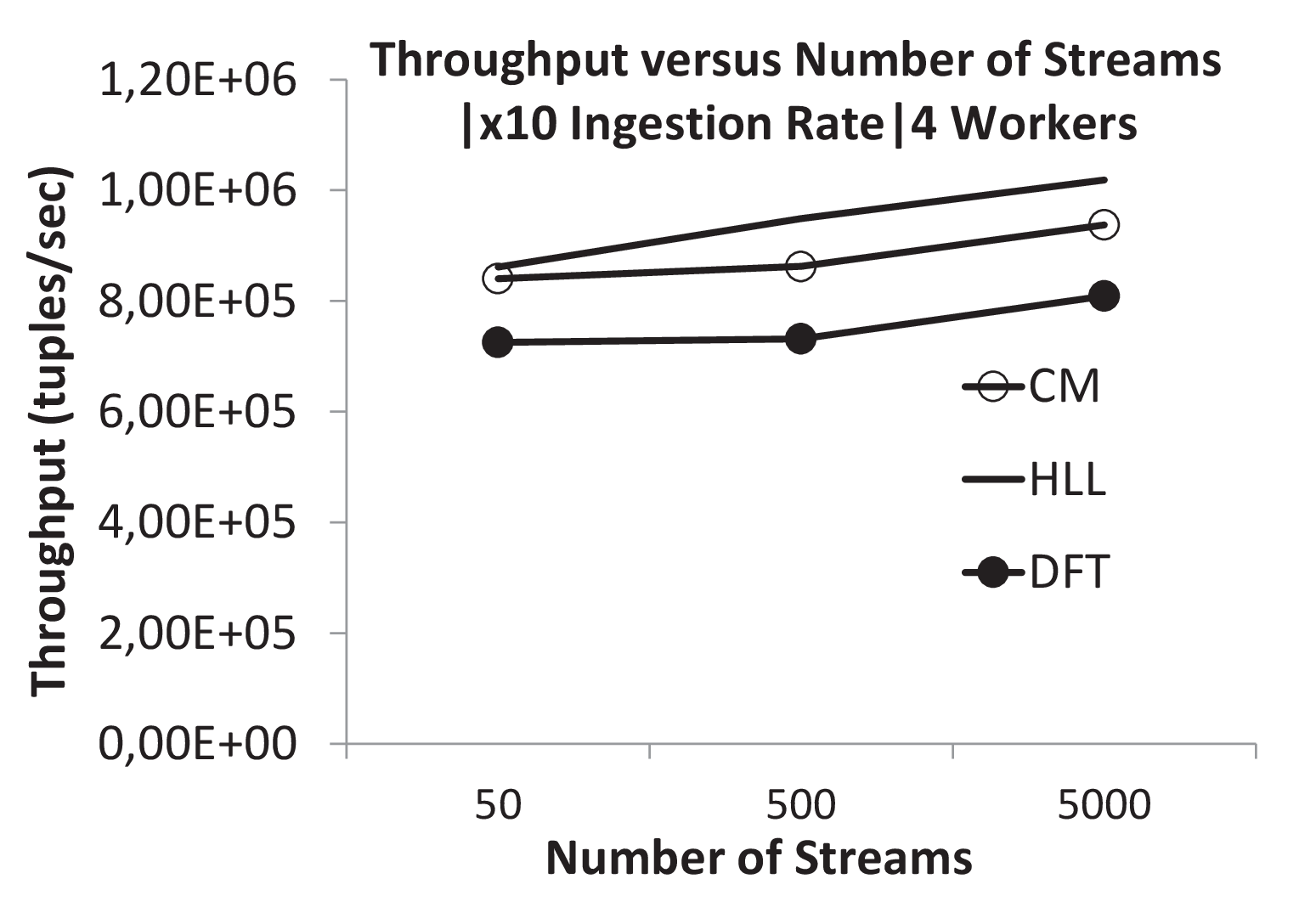}}
\subfigure[Varying the Number of Sites.]{\label{fig:clusters}\includegraphics[width=.245\textwidth ]{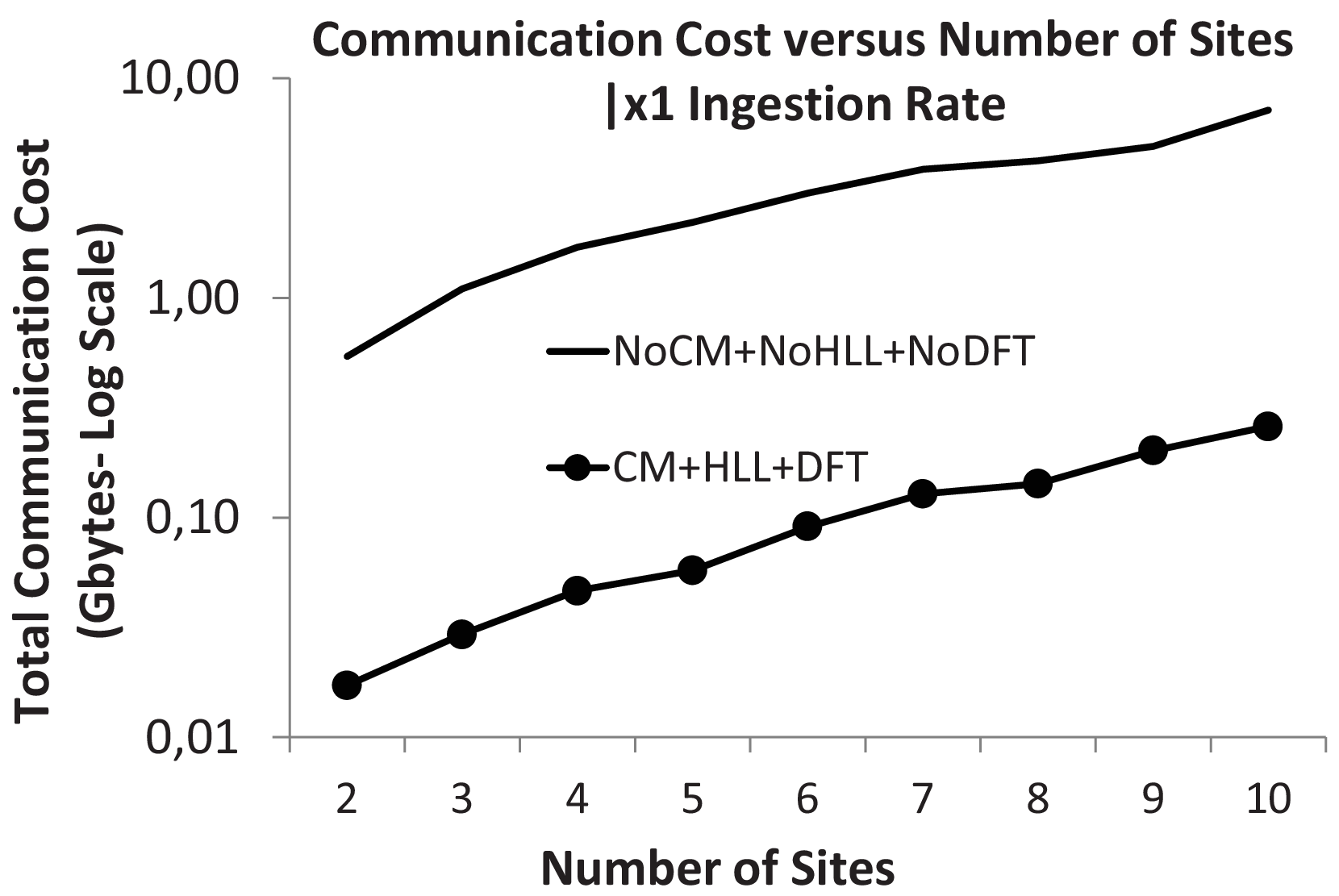}}
\caption{SDEaaS Scalability Study.}
\label{fig:scalability}
\end{figure*}

To test the performance of our SDEaaS approach, we utilize a Kafka cluster with 3 Dell PowerEdge R320 Intel Xeon E5-2430 v2 2.50GHz machines with 32GB RAM each and
one Dell PowerEdge R310 Quad Core Xeon X3440 2.53GHz machine with 16GB RAM. Our Flink cluster has 10 
Dell PowerEdge R300 Quad Core Xeon X3323 2.5GHz machines with 8GB RAM each.
We use a real dataset composed of $\sim$5000 stocks contributing a total of $\sim$10 TB of Level 1 and Level 2 data
provided to us by http://www.springtechno.com/ in the scope of the EU H2020 INFORE project (http://infore-project.eu/) acknowledged in this work. Note that our experiments concentrate on computational and communication performance figures. We do not
provide results for the synopses accuracy, since our SDEaaS
approach does not alter in anyway the accuracy guarantees of synopses. Theoretic bounds and experimental results for the accuracy of each synopsis can be found in
related works cited in Table~\ref{tab:library}.

\subsection{Assessing Scalability}
In the experiments of this first set, we test the performance of our SDEaaS approach alone.
That is we purely measure its performance on
maintaining various types of synopses operators, without placing these operators provided by the SDE as parts of a workflow.
In particular, we measure the throughput, expressed as the number
of tuples being processed per time unit (second) and communication cost (Gbytes) among workers, while varying a number of parameters involving horizontal ((i),(ii)), vertical (iii) and federated (iv) scalability, respectively: (i)
the parallelization degree [2-4-6-8-10], (ii)
the update ingestion rate [1-2-5-10] times the Kafka ingestion rate (i.e., each tuple read from Kafka is cloned [1-2-5-10] times in memory to further increase the tuples to process),
(iii) the number of summarized stocks (streams) [50-500-5000]  and
(iv) the Gbytes communicated among workers for maintaining each examined synopsis as a federated one. Note that this also represents the communication
cost that would incur among equivalent number of sites (computer clusters), instead of workers, each of which maintains its own synopses.
In each experiment of this set, we build and maintain Discrete Fourier Transform (DFT--8 coefficients, 0.9 threshold), HyperLogLog (HLL -- 64 bits, $m=3$), CountMin (CM -- $\epsilon=0.002,\delta=0.01$), AMS ($\epsilon=0.002, \delta=0.01$)
synopses each of which, as discussed in Section~\ref{sec:workflows}, is destined to support different types of analytics related to correlation, distinct count
and frequency estimation, respectively (Table~\ref{tab:library}). Since the CM and the AMS sketches exhibited very similar performance
we only include CM sketches in the graph to improve readability. All the above parameters were set after discussions with experts from the data provider and on the same ground, we use a time window of 5 minutes.  

Figure~\ref{fig:parallelism} shows that increasing the number of Flink workers causes proportional increase
in throughput. This comes as no surprise, since for steady ingestion
rate and constant number of monitored streams, increasing the parallelization degree causes fewer streams
to be processed per worker which in turn results in reduced processing load for each of them.
Figure~\ref{fig:injection}, on the other hand, shows that varying the ingestion rate from 1 to 10
causes throughput to increase almost linearly as well. This is a key sign of horizontal scalability, since
the figure essentially says that the data rates the SDEaaS can serve, quantified in terms of throughput,
are equivalent to the increasing rates at which data arrive to it. Figure~\ref{fig:sources} shows something
similar as the throughput
increases upon increasing the number of processed streams from 50 to 5000. This validates our
claim regarding the vertical scalability aspects the SDEaaS can bring in the workflows it participates. We further comment on
such aspects in the comparative analysis in Section~\ref{sec:comparisons}.

Finally,
Figure~\ref{fig:clusters} illustrates the communication performance of SDEaaS upon maintaining federated synopses and communicating
the results to a responsible site so as to derive the final estimations (see yellow arrows in Figure~\ref{fig:architecture}
and Section~\ref{sec:components}). For this experiment, we divide the streams among workers and each worker represents a site which analyzes its own stocks by computing CM, HLL, DFT synopses. A random site is set responsible for merging partial, local summaries and for providing the overall
estimation, while we measure the total Gbytes that are communicated among sites/workers as more sites along with their
streams are taken into consideration. Note that the sites
do not communicate all the time, but upon an \texttt{Ad-hoc Query} request every 5 minutes.  

Here, the total communication cost for deriving estimations from synopses, is not a number that says much on its own. It is
expected of the communication cost will rise as more sites are added to the network. The important factor to judge
federated scalability is the communication cost when we use the synopses ("CM+HLL+DFT" line in Figure~\ref{fig:clusters}) compared to when we do not. Therefore, in Figure~\ref{fig:clusters}, we also plot a line (labeled "NoCM+NoHLL+NoDFT") illustrating the
communication cost that takes place upon answering the same (cardinality, count, time series) queries without
synopses. As Figure~\ref{fig:clusters} illustrates (the vertical axis is in log scale),
the communication gains steadily remain above an order of magnitude.

\subsection{Comparison against other Candidates}\label{sec:comparisons}
 We use the DFT synopsis to replace \texttt{Window}, \texttt{AggregativeOperation} as discussed in Section~\ref{sec:workflows}, since the most computationally intensive (and thus candidate to become the bottleneck) operator in the workflow
of Figure~\ref{fig:workflow} is the \texttt{AggregativeOperation} which performs pairwise correlation estimations of time series.
Indicatively, when 5K stocks are monitored, the pairwise similarity comparisons that need to be performed by naive approaches are
12.5M.

\begin{figure}[t]
\centering \includegraphics[width=.6\columnwidth]{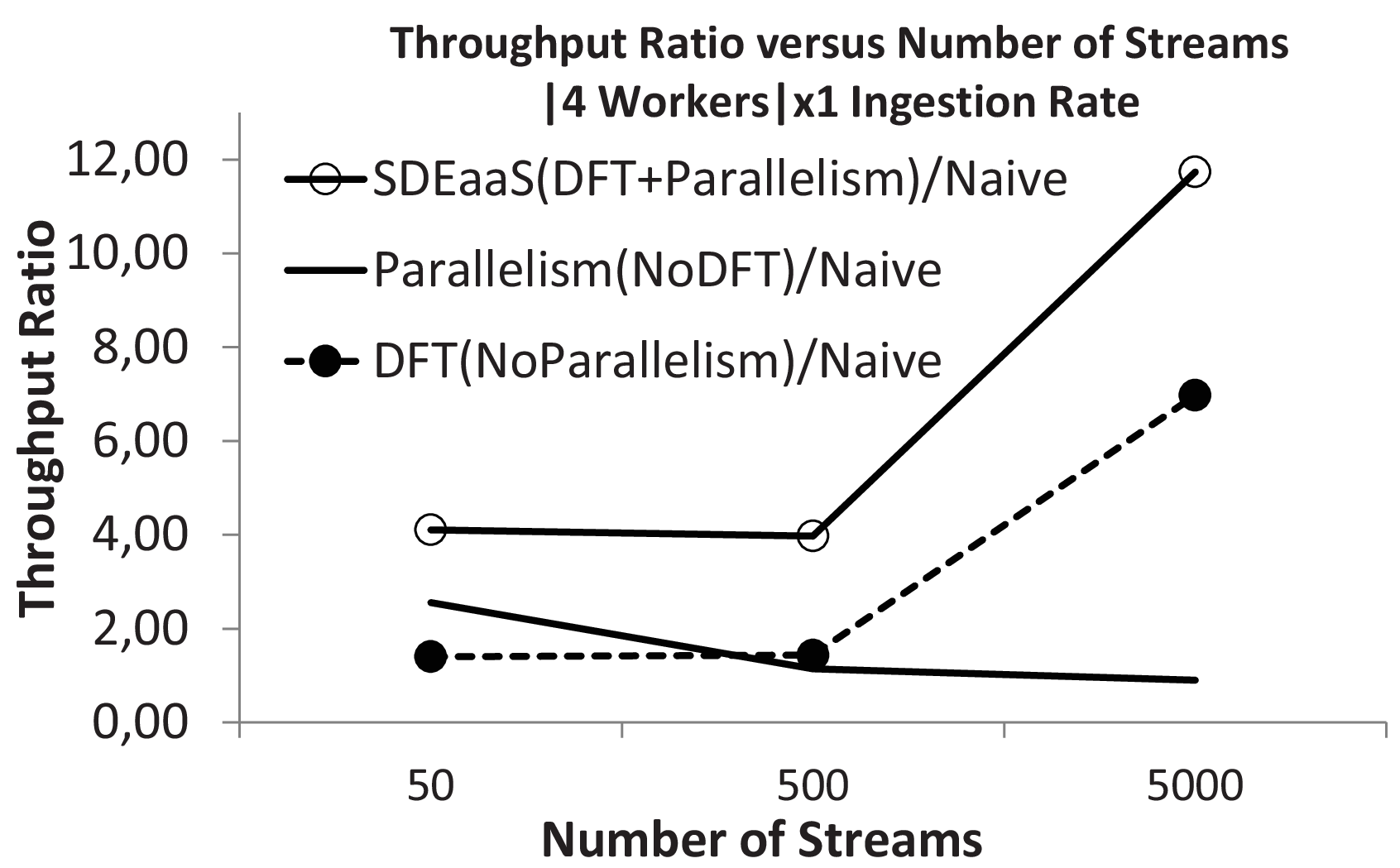}
\caption{Comparative Analysis in Executing the Workflow of Figure~\ref{fig:workflow} using \texttt{SDE.DFT}.}
\label{fig:comparison}
\end{figure}

In Figure~\ref{fig:comparison} we measure the performance of our SDEaaS approach employed in this work against three
alternative approaches. More precisely, the compared approaches are:
\begin{itemize}[leftmargin=*]
\item \noindent{\bf Naive\/}: This is the baseline approach which involves sequential processing of incoming tuples without parallelism or any synopsis.

\item \noindent{\bf SDEaaS(DFT+Parallelism)\/}: This is the approach employed in this work which combines the virtues of parallel processing (using 4 workers in Figure~\ref{fig:comparison}) and stream summarization (DFT synopsis) towards delivering interactive analytics at extreme scale. 

\item \noindent{\bf Parallelism(NoDFT)\/}: This approach performs parallel processing (4 workers), but does not utilize any
synopses to bucketize time series or reduce their dimensionality. Its performance corresponds to competitors such as SnappyData~\cite{DBLP:reference/bdt/Mozafari19} which provide
an SDE, but their SDE is restricted to simple aggregates, thus neglecting synopses ensuring vertical scalability. Moreover, for the same reason, it also represents the performance of synopsis utilities provided by Spark. 

\item \noindent{\bf DFT(NoParallelism)\/}: The DFT(NoParallelism) approach utilizes DFT synopses to bucketize time series and for dimensionality reduction, but no parallelism is used for executing the workflow of Figure~\ref{fig:workflow}. Pairwise similarity checks are restricted to adjuscent buckets and thus comparisons can be pruned, but the computation of similarities is not performed in parallel for each bucket. This approach corresponds to competitors such as DataSketch~\cite{DataSketch} or Stream-lib~\cite{stream-lib} which provide a synopses library but do
not include parallel implementations of the respective algorithms and do not follow an SDEaaS paradigm. 
\end{itemize}

Each line in the plot of Figure~\ref{fig:comparison} measures the ratio of throughputs of each examined approach over the Naive approach
varying the amount of monitored stock streams. 
Let us first examine each line individually. It is clear that when we monitor few tens of stocks (50 in the figure), the use of DFT in the
DFT(NoParallelism) marginally improves (1.5 times higher throughput) the throughput of the Naive approach. On the other hand, the Parallelism(NoDFT) improves
the Naive by $\sim$2.5 times. Our SDEaaS(DFT+Parallelism), taking advantage of both the synopsis and parallelism improves the Naive by
almost 4 times. Note that when 50 streams are monitored, the number of performed pair-wise similarity checks in the workflow of
Figure~\ref{fig:workflow} for the Naive approach is 2.5K/2. 

This is important because, according to Figure~\ref{fig:comparison}, when we switch to
monitoring 500 streams, i.e., 250K/2 similarity checks are performed by Naive, the fact that the Parallelism(NoDFT) approach lacks the
ability of the DFT to bucketize time series and prune unnecessary similarity checks, makes its throughput approaching the Naive
approach. This is due to \texttt{AggregativeOperation} starting to become
a computational bottleneck for Parallelism(NoDFT) in the workflow of Figure~\ref{fig:workflow}. On the contrary, the DFT(NoParallelism) line remains steady when switching from 50 to 500 streams. The DFT(NoParallelism) approach starts to perform better than Parallelism(NoDFT) on 500 monitored streams showing that the importance of comparison pruning and, thus, of vertical scalability is higher than the importance of parallelism, as more streams are monitored. The line corresponding to our SDEaaS(DFT+Parallelism) approach exhibits steady behavior upon switching from 50 to 500, improving the Naive approach by 4 times, the DFT(NoParallelism) approach by 3
and the Parallelism(NoDFT) approach by 3.5 times. 

The most important findings come upon switching to monitoring 5000 stocks
(25M/2 similarity checks using Naive or Parallelism(NoDFT)).
Figure~\ref{fig:comparison} says that because of the lack of the vertical scalability provided by the DFT, the Parallelism(NoDFT) approach becomes equivalent to the Naive one. The DFT(NoParallelism) approach improves the throughput of the Naive and
of Parallelism (NoDFT) by 7 times. Our SDEaaS(DFT+Parallelism) exhibits 11.5 times better performance compared to Naive, Parallelism(NoDFT) and almost doubles the performance of DFT(NoParallelism). This validates the potential of SDEaaS(DFT+Parallelism) to support
interactive analytics upon judging similarities of millions of pairs of stocks. In addition, studying the difference between
DFT(NoParalleli-sm) and SDEaaS(DFT+Parallelism) we can quantify which part of the improvement over Naive, Parallelism(NoDFT) is caused due to comparison pruning
based on time series bucketization and which part is yielded by parallelism. That is, the use of DFT for bucketization and
dimensionality reduction increases throughput by 7 times (equivalent to the performance of DFT(NoParallelism)), while
the additional improvement entailed by SDEaaS(DFT+Parallelism) is roughly equivalent to the number of workers (4 workers in Figure~\ref{fig:comparison}). This indicates the success of SDEaaS in integrating the virtues of data synopsis and parallel processing.

\begin{figure}[t]
\centering \includegraphics[width=.6\columnwidth]{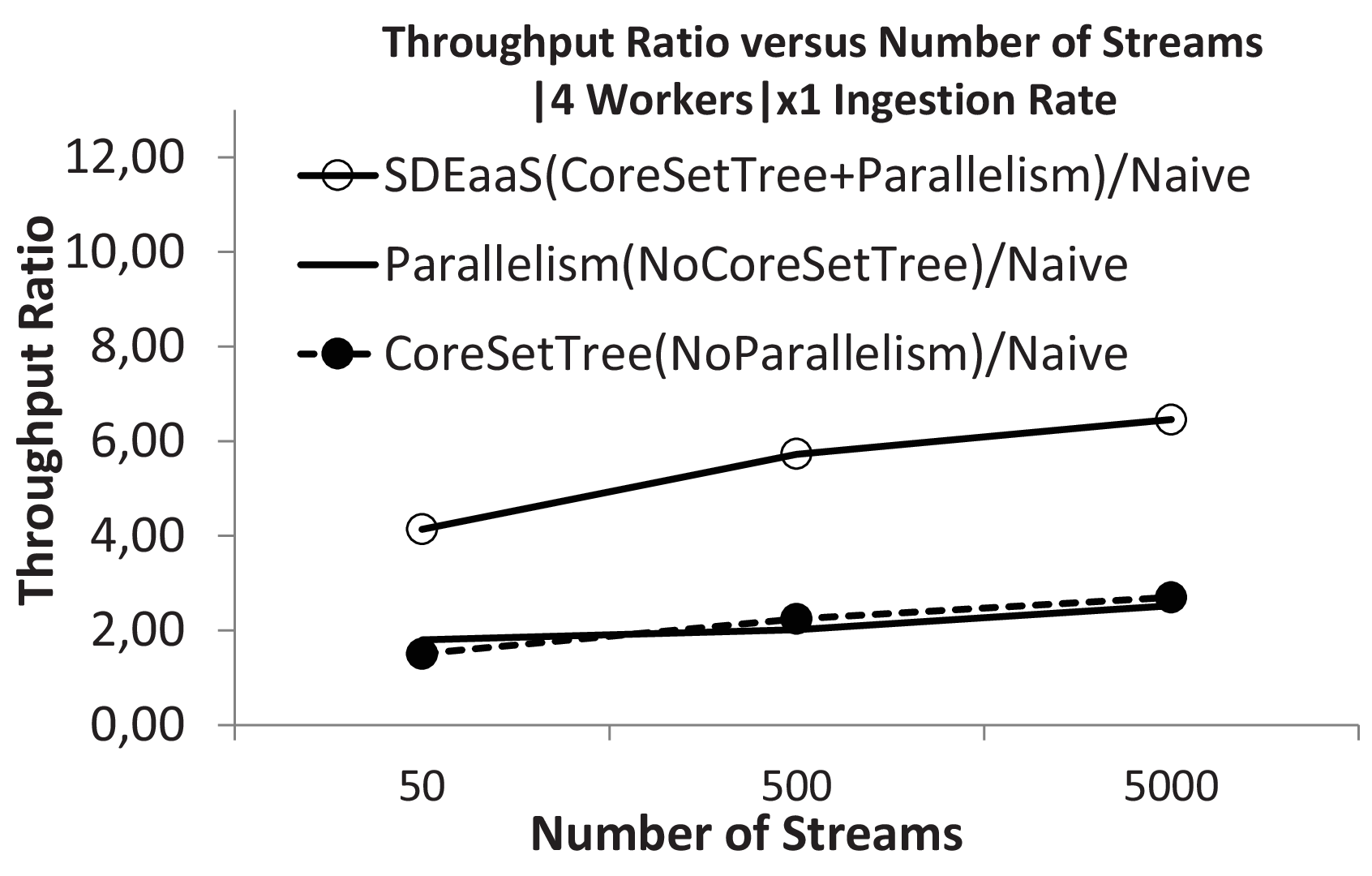}
\caption{Comparative Analysis in Executing the Workflow of Figure~\ref{fig:workflow} using \texttt{SDE.CoreSetTree}.}
\label{fig:comparison2}
\end{figure}

We then perform a similar experiment for the stream mining version of the workflow in Figure~\ref{fig:workflow} as described
in Section~\ref{sec:workflows}. In particular,
in this experiment the Naive approach corresponds to StreamKM++ clustering without parallelism and coreset sizes equivalent to the original data points (time series). The Parallelism(NoCoreSetTree) approach involves
performing StreamKM++ with coreset sizes equivalent to the original data points, but exploiting parallelism. The CoreSetTree(NoParallelism)
exploits the CoreSetTree synopsis but uses no parallelism, while SDEaaS(CoreSetTree +Parallelism) combines the two.
For CoreSetTree(NoParallelism) and SDEaaS(CoreSetTree+Parallelism), we use bucket sizes of 10-100-400 and $k$
values are set to $4-10-40$, for 50-500-5000 streams, correspondingly. The conclusions that can be drawn
from Figure~\ref{fig:comparison2} are very similar
with what we discussed in Figure~\ref{fig:comparison}. However, the respective ratios of throughput over the Naive approach
are lower (2-3 times higher throughput than the second best candidate in Figure~\ref{fig:comparison2}). This is by design of the mining algorithm and the reason is that the clustering procedure includes
a reduction step which is performed by a single worker. This is in contrast with the \texttt{ApplyThreshold} operation in
Figure~\ref{fig:comparison} which can be performed by different processing units independently.  

\subsection{SDEaaS vs non-SDEaaS}

In Section~\ref{sec:lessons} we argued about the fact that employing a non-SDEaaS approach, as works such as~\cite{proteus} do,
restricts the maximum allowed number of concurrently maintained synopses up to the available
task slots. That is, if the SDE is not provided as a service using our novel architecture, in case we want to maintain a
new synopsis when a demand arises (without ceasing the currently maintained ones, because these may
already serve workflows as the one in Figure~\ref{fig:workflow}), we have to submit a new job. A job
occupies at least one task slot. On the contrary, in our SDEaaS approach, when a request for a new synopsis
arrives on-the-fly, we simply devote more tasks (which can exploit hyper-threading, pseudo-parallelism etc)
instead of entire task slots. Because of that, our SDEaaS design is a much more preferable choice since it
can simultaneously maintain thousands of synopses for thousands of streams. 

\begin{figure}[h]
\centering \includegraphics[width=.6\columnwidth]{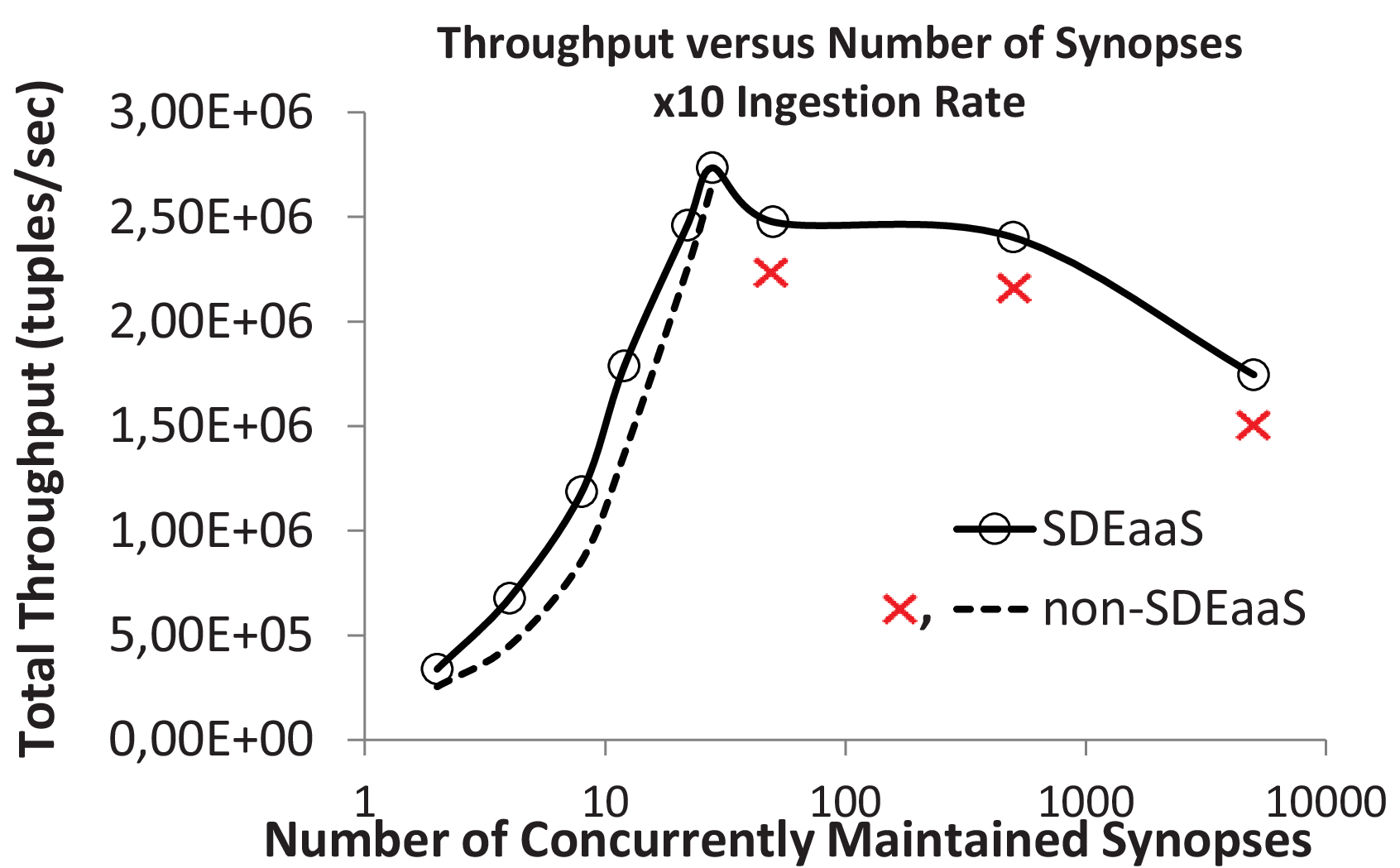}
\caption{Comparison of SDAaaS vs non-SDEaaS. \textcolor{red}{\ding{56}} signs denote that
non-SDEaaS cannot maintain more than 40 synopses simultaneously since available task slots are depleted.}
\label{fig:vsnonSDEaaS}
\end{figure}

To show the superiority of our approach in practice, we design an experiment where we start with
maintaining 2 CM sketches for frequency estimations on the volume, price pairs of each stock.
Note that this differs compared to what we did in Figure~\ref{fig:scalability} where we kept a
CM sketch for estimating the count of bids per stock in the whole dataset.  
Then, we express demands for maintaining one more CM sketch for up to 5000 sketches/stocks. We do that
without stopping the already running synopses each time. We measure the sum of throughputs of all running
jobs for the non-SDEaaS approach and the throughput of our SDE and plot the results in Figure~\ref{fig:vsnonSDEaaS}.

First, it can be observed that we cannot maintain more than 40 synopses simultaneously using the non-SDEaaS approach
since we deplete the available task slots. This is denoted with \textcolor{red}{\ding{56}} signs in the plot. Second, even when up to 40 synopses
are concurrently maintained, our SDEaaS approach always performs better compared to the non-SDEaaS alternative.
This is because slot sharing in SDEaaS means that more than one task is scheduled into the same slot, or in other words, CM sketches end 
up sharing resources. The main benefit of this is better resource utilization. In the non-SDEaaS approach if there is skew in the update 
rate of a number of streams (to which one task slot per synopsis per stream is alloted), we might easily end 
up with some slots doing very little work at certain intervals, while others are quite busy. This is avoided in SDEaaS due to slot sharing. Therefore, better resource utilization is an additional advantage of our SDEaaS approach.

\section{Conclusions and Future Work}

In this work we introduced a Synopses Data Engine (SDE) for enabling interactive analytics
over voluminous, high-speed data streams. Our SDE is implemented following a SDE-as-a-Service (SDEaaS) paradigm and is materialized via
a novel architecture. It is easily extensible, customizable with new synopses and capable of providing various types of
scalability. Moreover, we exhibited ways in which SDEaaS can serve workflows for different purposes and we commented on
implementation insights and lessons learned throughout this endeavor. Our future work focuses
on (a) enriching the SDE Library with more synopsis techniques~\cite{DBLP:journals/ftdb/CormodeGHJ12}, (b) 
integrate it with machine learning components such as~\cite{proteus}, (c) implement the proposed SDEaaS
architecture directly on the data ingestion layer via Kafka Streams which lacks facilities like \texttt{CoFlatMap},
(d) similarly for Apache Beam~\cite{beam}, to make the service directly runnable to a variety of Big Data platforms.

\bibliographystyle{ACM-Reference-Format}

%
% ...

\end{document}